\documentclass[msom,nonblindrev]{style/informs3}
\usepackage{booktabs} 
\usepackage{subcaption}
\usepackage{accents}

\OneAndAHalfSpacedXI 

\usepackage{natbib}
 \bibpunct[, ]{(}{)}{,}{a}{}{,}%
 \def\bibfont{\small}%

\AtBeginDocument{

}%

\usepackage{threeparttable}
\usepackage{calc}
\usepackage{enumitem,linegoal}
\usepackage[hidelinks]{hyperref}
\usepackage{float}
\usepackage{soul}
\usepackage{tabularx}
\usepackage{subcaption,booktabs}
\usepackage[linesnumbered,ruled,vlined]{algorithm2e}
\usepackage[normalem]{ulem}
\usepackage{url}
\usepackage{amsmath}
\usepackage{graphicx}
\usepackage{subcaption}
\usepackage{dsfont}
\usepackage{xcolor}
\usepackage{xspace}

\AtBeginDocument{

}

\TheoremsNumberedThrough     
\ECRepeatTheorems

\EquationsNumberedThrough    

\MANUSCRIPTNO{}

\newcommand{\aref}[1]{\hyperref[#1]{Appendix~\ref*{#1}}}

\begin{document}



\RUNTITLE{AI Self-preferencing in Algorithmic Hiring: Empirical Evidence and Insights}

\TITLE{AI Self-preferencing in Algorithmic Hiring: Empirical Evidence and Insights}

\ARTICLEAUTHORS{%
\AUTHOR{Jiannan Xu}
\AFF{Robert H. Smith School of Business, University of Maryland, MD 20742, United States, \\ 
\EMAIL{jiannan@umd.edu}}
\AUTHOR{Gujie Li}
\AFF{School of Computing, National University of Singapore, 
Singapore 117417, \\ \EMAIL{gujieli@nus.edu.sg}}
\AUTHOR{Jane Yi Jiang}
\AFF{Max M. Fisher College of Business, The Ohio State University, OH 43210, United States, \\ \EMAIL{jiang.3186@osu.edu}}
}

\ABSTRACT{\textbf{\emph{Problem definition}}: As artificial intelligence (AI) tools become widely adopted, large language models (LLMs) are increasingly involved on both sides of decision-making processes, ranging from hiring to content moderation. This dual adoption raises a critical question: do LLMs systematically favor content that resembles their own outputs? Prior research in computer science has identified self-preference bias, i.e., the tendency of LLMs to favor their own generated content, but its real-world implications have not been empirically evaluated. 
\textbf{\emph{Methodology/results}}: We examine this question in the context of algorithmic hiring, a high-volume screening process central to workplace operations. In this setting, job applicants increasingly rely on LLMs to refine resumes, while employers deploy similar tools to screen those same materials. Using a large-scale controlled resume correspondence experiment, we find that LLMs consistently prefer resumes generated by themselves over those written by humans or produced by alternative models, even when content quality is controlled. The bias against human-written resumes is particularly substantial, with self-preference bias ranging from $67\%$ to $82\%$ across major commercial and open-source models. 
\textbf{\emph{Managerial implications}}: To quantify the operational impact of self-preference bias, we simulate realistic hiring pipelines across $24$ occupations. These simulations show that candidates using the same LLM as the evaluator are $23\%$ to $60\%$ more likely to be shortlisted than equally qualified applicants submitting human-written resumes, with the largest disadvantages observed in business-related fields such as sales and accounting. We further demonstrate that, in many cases, this bias can be reduced by more than $50\%$ through simple interventions that target LLMs' self-recognition capabilities. These findings highlight an emerging but previously overlooked risk in AI-assisted decision making and call for expanded frameworks of AI fairness in business operations that address biases from AI-AI interactions.
}%


 \HISTORY{This paper was written on August 29th, 2025.}
\maketitle
%


\section{Introduction}
The rapid development and commercialization of artificial intelligence (AI) have made large language models (LLMs) widely accessible across both professional and everyday contexts. As these tools become embedded in diverse workflows, they are increasingly involved on both sides of content generation and evaluation. In hiring, for instance, applicants often use LLMs to draft or refine their resumes, while employers leverage similar tools to screen or rank candidates \citep{NYSSCPA2024AIHiring, ResumeBuilder2023ChatGPTResume, wiles2025algorithmic}. On social media, users frequently use LLMs to help compose or polish posts, while platforms may employ LLMs to moderate content, e.g., flagging, categorizing, or filtering user submissions (e.g., \citealt{kumar2024watch}). In academia, researchers may use LLMs to improve their manuscripts, while conferences and journals are beginning to experiment with LLM-assisted peer review (e.g., \citealt{thakkar2025can}).  Similar patterns are emerging in education and customer service, where LLMs support both communication and assessment tasks \citep{forbes_llms_customer_support_2024}.
In these domains and beyond, LLMs increasingly \emph{both} generate and evaluate the same content, giving rise to a new class of AI–AI interactions with significant implications for human decision-making and business operations \citep{laurito2025ai}.

Recent research in computer science has identified a behavioral tendency in LLMs known as self-preference, i.e., the inclination of a model to favor content it generated itself over that written by humans or produced by alternative models (e.g., \citealt{Panickssery2024}). While this phenomenon has been documented in benchmark evaluations (e.g., \citealt{zheng2023judging, Bai2023}), its implications for real-world decision-making remain largely unexplored. As LLMs are increasingly deployed in high-stakes settings where they may evaluate content also generated by LLMs, self-preference introduces a novel form of bias. Unlike traditional biases rooted in demographic disparities (e.g., \citealt{sheng2019woman}), this bias emerges endogenously from AI-AI interactions, in which the model's own evaluative behavior systematically favors outputs aligned with its generative patterns.

Self-preferencing bias is inherently interactional: it arises when LLMs are asked to judge content that may share stylistic or linguistic patterns with their own generative outputs. As such, it poses a new challenge for AI fairness, one that is not addressed by existing safeguards focused on demographic disparities. If left unchecked, self-preference could subtly distort evaluative processes across hiring, education, publishing, and more---privileging those who employ the same AI system used for evaluation (i.e., the ``right'' tool from the model's perspective) while disadvantaging those who use different tools or none at all. Addressing this issue will require expanding current fairness frameworks to account for LLMs' dual roles as both decision aids and evaluators in an increasingly AI-mediated business environment. 

Among the many domains where self-preference bias may arise, algorithmic hiring is particularly consequential. Employers are now routinely adopting LLMs to streamline resume screening and candidate ranking, often as part of automated workflows that support human decision-making (e.g., \citealt{gan2024application, kim2024llm, resumebuilder2025aihiring, sarumathi2025ai}). Unlike traditional keyword-matching systems, LLMs can evaluate resumes in a more holistic manner—synthesizing content, inferring intent, and making contextual judgments beyond simple heuristics \citep{Pritchett2025_AIresume}. While this shift promises greater efficiency and scalability, it may also magnify the risk of bias if an LLM systematically favors resumes that reflect its own generative style. In such cases, evaluations may hinge less on the substantive quality of a candidate's credentials and more by superficial stylistic alignment with the evaluator LLM, conferring unwarranted advantages on applicants who use the same model to compose their materials. In effect, this bias rewards access to specific generative technologies and penalizes those without it, even when applicants are otherwise equally qualified. 

In this paper, we provide the first empirical evidence that self-preference bias can distort candidate evaluations in algorithmic hiring. Specifically, we examine whether LLMs, when deployed as evaluators, systematically favor resumes they generated themselves over otherwise equivalent resumes written by humans or produced by alternative LLM models. To test this, we conduct a large-scale resume correspondence experiment using a real-world dataset of $2,245$ human-written resumes, sourced from a professional resume-building platform prior to the widespread adoption of generative AI. For each resume, we generate multiple counterfactual versions using a range of state-of-the-art LLMs, including GPT-4o, GPT-4o-mini, GPT-4-turbo, LLaMA 3.3-70B, Mistral-7B, Qwen 2.5-72B, and Deepseek-V3.\footnote{These models represent the state of the art among widely deployed LLMs at the time of our study. Although newer and more capable models will continue to emerge, the key pattern we examine---LLMs being used both to generate content and to evaluate that content---already appears in many application domains and is expected to remain common. Our focus is therefore on a structural feature of AI-mediated decision making rather than on the behavior of any specific model version.} Having content quality controlled, we assess whether these LLMs exhibit systematic bias in favor of their own outputs when acting as evaluators. 

We distinguish between two forms of self-preference bias: \emph{LLM-vs-Human}, where a model prefers its own generated content over a human-written equivalent; and \emph{LLM-vs-LLM}, where a model favors its own output over content produced by an alternative (different) LLM. We find strong and consistent evidence of LLM-vs-Human self-preferencing across most models. Larger models, such as GPT-4o, GPT-4-turbo, DeepSeek-V3, Qwen-2.5-72B, and LLaMA 3.3-70B, exhibit particularly strong bias, exceeding $65\%$ even after controlling for content quality and reaching over $80\%$ for GPT-4o. By contrast, LLM-vs-LLM self-preferencing is more heterogeneous. DeepSeek-V3 shows the strongest bias in this setting, preferring its own outputs by $69\%$ against LLaMA 3.3-70B and $28\%$ against GPT-4o. GPT-4o and LLaMA 3.3-70B, in comparison, do not display consistent preferences when evaluating content generated by other models.


To assess the labor market implications of this bias, we simulate hiring pipelines across $24$ occupations. In these simulations, candidates using the same LLM as the evaluator are about $23$–$60\%$ more likely to be shortlisted than equally qualified applicants submitting human-written resumes. The disadvantage is most severe in business-related fields such as accounting, sales, and finance, and less pronounced in areas like agriculture, arts, and automotive. If similar advantages were to persist across repeated hiring cycles, they could give rise to a ``lock-in'' effect, in which the stylistic patterns favored by dominant LLMs gradually become entrenched in applicant pools, with the potential to amplify inequities and reduce diversity in candidate selection. 

To mitigate AI self-preference, we propose two simple yet effective strategies that directly target the underlying mechanism of self-recognition---a model's ability to implicitly identify content it generated. The first strategy uses system prompting to explicitly instruct models to ignore the origin of resumes and focus only on substantive content. The second strategy employs a majority voting ensemble, combining the evaluator model with smaller models that exhibit weaker self-recognition, thereby diluting the bias of any single LLM. Across all tested LLMs, these interventions reduce LLM-versus-human self-preference by $17\%$ to $63\%$ in relative terms. These results demonstrate that while self-preference bias is widespread and consequential, it is not immutable: straightforward design interventions can substantially improve fairness in LLM-based hiring operations.

Together, these findings advance our understanding of how generative AI systems shape operational decision processes in organizations. We document and quantify self-preference bias in the context of resume screening by developing a measurement framework grounded in established fairness metrics. In addition, we show that self-preference bias can be substantially reduced through targeted interventions informed by LLM's self-recognition behavior. In doing so, the study introduces a novel and practical perspective on AI fairness for business operations---one that moves beyond concerns about demographic disparities to address interactional biases that arise when AI systems evaluate content they themselves could have produced.

The remainder of the paper is organized as follows. We begin by reviewing the related literature in \autoref{section: literature}. \autoref{section: methodology} defines and outlines the measurement of AI self-preferencing bias. \autoref{section: data} describes the dataset and experimental design, and \autoref{section: empirical_results} presents the empirical findings. In \autoref{sec:mitigation}, we introduce and evaluate mitigation strategies. Finally, we conclude in \autoref{section: concluding_remarks} with a discussion of key implications and directions for future research.

\section{Literature Review}
\label{section: literature}
Our research contributes to multiple strands of literature: (1) screening and hiring as operational processes, (2) generative AI in labor market and workplace operations, (3) algorithmic fairness, bias, and responsible operations, and (4) LLMs as judges in decision making. 



\subsection{Screening and Hiring as Operational Processes}
Operations Management (OM) research has long examined hiring and screening as core operational processes through which organizations allocate scarce evaluation capacity and form productive matches under uncertainty (e.g., \citealt{freeman1983secretary, jovanovic1979job, arlotto2014optimal, yoo2016time, koren2026gatekeeper}). This work positions workforce selection at the interface between operations and human resource management \citep{boudreau2003interface}, conceptualizing recruitment and screening not as purely administrative functions but as operational processes through which firms transform applicant pools into productive labor inputs that support day-to-day operations. 

A central implication of this lens is that hiring pipelines are typically \emph{multi-stage} and \emph{capacity constrained}. Upstream screening (e.g., resume review) funnels a large applicant pool into a limited set of downstream evaluations (e.g., interviews and assessments), where time, managerial attention, and specialist resources are scarce. As in other service and information-processing systems, errors at upstream triage have asymmetric and persistent consequences: false negatives irreversibly eliminate qualified candidates, while false positives consume scarce evaluation capacity and degrade downstream match quality. Recent work in platform and labor market operations further emphasizes how design choices in these screening stages shape congestion, equilibrium outcomes, and match efficiency. For example, \citet{farajollahzadeh2025rooney} show that interview-stage diversity interventions such as the Rooney Rule can backfire in capacity-constrained settings with noisy pre-interview signals, while \citet{baek2025hiring} demonstrate that strategic interview decisions in multistage, multi-firm hiring processes can substantially mitigate market inefficiencies and increase the number of hires overall.

Our paper contributes to this stream of literature by identifying a new and operationally consequential distortion that arises at the \emph{earliest} stage of the hiring funnel when screening is delegated to AI evaluators. Existing OM models implicitly assume that evaluators process signals in a manner that is neutral to how those signals are generated. We relax this assumption and offer empirical evidence by studying an environment in which applicants increasingly use LLMs to generate screening inputs (e.g., resumes), while firms deploy similar models to evaluate those same inputs. We show that AI evaluators can exhibit systematic self-preference, favoring content generated by the same model, even when underlying candidate information is held constant. This interaction-driven bias is distinct from standard screening noise: it is endogenous to pipeline design choices (which tools are used upstream and downstream) and can create path-dependent lock-in of ``dominant'' resume styles, with implications for both operational performance (misranking and misallocation) and responsible hiring system design.

\subsection{Generative AI in Labor Market and Workplace Operations}
A growing literature examines the role of AI in labor market and workplace operations. Research spanning operations management, economics, and information systems documents how AI tools are increasingly deployed in both production and evaluation tasks, with important implications for worker performance, learning, and market outcomes. For example, \citet{brynjolfsson2025generative} provide field evidence that a generative AI assistant improves productivity in customer support, with heterogeneous effects depending on workers' experience and baseline skill. In labor-market contexts, a complementary stream studies how AI-assisted writing affects the formation of applicant signals and subsequent hiring outcomes. \citet{wiles2025algorithmic} show that algorithmic resume writing assistance improves writing quality and increases hiring, suggesting that generative AI can meaningfully alter the informational content and effectiveness of applicant signals in matching markets. At the same time, the introduction of algorithms into evaluation processes raises well-documented concerns. Prior work shows that human decision makers may discount or resist algorithmic recommendations (e.g., \citealt{dietvorst2015algorithm}), and that algorithmic hiring systems can encode or amplify existing biases when trained on historical data or used with discretion (e.g., \citealt{hoffman2018discretion, cowgill2018bias}). 

Our paper contributes to this emerging body of work by identifying a distinct and previously unexamined mechanism that arises when generative AI is used on both sides of the hiring market. We provide controlled, large-scale evidence that such coupling can create systematic advantages for certain applicants even when quality is held constant. Specifically, we document and quantify AI self-preferencing in resume screening. We then connect this bias to labor-market consequences by simulating capacity-constrained shortlisting pipelines across occupations, showing how evaluator self-preference affects the allocation of interview opportunities. Finally, we demonstrate that simple interventions (e.g., system prompting and majority-vote ensemble) can substantially mitigate this distortion, offering practical guidance for firms adopting generative AI in hiring workflows.

\subsection{Algorithmic Fairness, Bias, and Responsible Operations}
Our work also contributes to the growing literature on responsible practices in operations management, which emphasizes that operational decisions should be evaluated not only by efficiency but also by their broader implications for fairness and stakeholder welfare. Recent research highlights that dimensions such as fairness, equity, and employee well-being are central---rather than peripheral---to the design of effective and sustainable operational systems (e.g., \citealt{lee2018socially, netessine2022om, kesavan2022doing, wang2023mind}). This body of work underscores a core insight: responsible operations require careful attention to how system design choices shape downstream human outcomes, often in context-specific ways.

Hiring and screening provide a particularly salient application of these ideas. Discrimination in labor markets has long been a central concern for both policymakers and economists, with extensive research documenting that certain race, gender, and age subpopulations face unfair treatment from recruiters in the hiring process (e.g., \citealt{bertrand2004emily, Kline2022, neumark2018experimental}). Concerningly, studies such as \citet{datta2015automated} and \citet{lambrecht2019algorithmic} have further shown that, beyond human bias, algorithms deployed in hiring systems can also encode discriminatory behavior that disadvantages particular demographic groups. In recent years, the rise of LLMs with their increasing adoption in hiring systems has prompted researchers to empirically evaluate whether emerging AI-driven hiring systems perpetuate similar forms of algorithmic discrimination. For instance, \citet{veldanda2023emily} replicated \citeauthor{bertrand2004emily}'s correspondence experiment to investigate AI-driven hiring bias across gender, race, maternity status, pregnancy status, and political affiliation. Additionally, \citet{an2024large} explored whether LLMs exhibit race- and gender-based discrimination through an experiment, in which LLMs write an email to a named job candidate about a hiring decision. Likewise, \citet{Nghiem2024} analyzed the extent to which LLMs exhibit bias toward applicants based on their first names when making employment recommendations.

At the same time, a broader interdisciplinary literature documents additional risks associated with the use of LLMs in organizational decision making, including hallucinations, systematic errors, and the reinforcement of social disparities (e.g., \citealt{Magesh2024, Huang2025}). These concerns have motivated extensive research on the ethical, legal, and governance implications of algorithmic hiring systems (e.g., \citealt{Raghavan2020, Li2021algo, Hunkenschroer2023}). For instance, \citet{Raghavan2020} study how firms that develop algorithmic pre-employment assessments build, validate, and manage bias in practice, highlighting gaps between technical capabilities, organizational incentives, and regulatory oversight. 

Despite this growing attention to algorithmic fairness and governance, existing work largely evaluates hiring algorithms in isolation and focuses on disparities across protected attributes. Unlike prior work that centers on protected characteristics, the self-preference bias we study arises from interactions among algorithmic components within the hiring pipeline itself. By uncovering this interaction-induced bias, our study highlights an emergent risk in algorithmic hiring systems that is not directly addressed by existing fairness audits or regulatory frameworks.

More broadly, our findings inform ongoing debates on AI governance and responsible operations by demonstrating how seemingly neutral design choices, such as deploying the same or similar models for content generation and evaluation, can systematically distort how scarce evaluation capacity is allocated across candidates. This perspective underscores the need for regulatory and organizational approaches that account not only for input data and protected attributes, but also for the interaction structure of AI-driven operational processes.




\subsection{LLMs as Judges in Decision Making}
The concept of LLM-as-a-Judge, first introduced by \citet{zheng2023judging}, refers to the use of LLMs as automated evaluators that assess responses and assign scores, and has gained traction in AI research as an efficient method for evaluating model performance without human intervention. However, emerging evidence suggests that LLM-based evaluation frameworks may introduce self-preference bias, where models disproportionately favor their own generated responses over those produced by humans or alternative models. For example, \citet{zheng2023judging} examine the potential of this paradigm but also identified its limitations---several biases, including self-enhancement bias (self-preference bias), arise when LLMs are used to judge responses, such as those they generate themselves. While the results show that strong models like GPT-4 achieve over $80\%$ agreement with human preferences on multi-turn conversations, they caution that LLM judges may still introduce systematic distortions, especially in high-stakes evaluation settings. Expanding on this line of inquiry, \citet{xu-etal-2024-pride} formally define self-preference bias and empirically revealed that it is widespread across popular LLMs and multiple tasks (e.g., translation, mathematical reasoning) through extensive experiments. They further demonstrate that the bias is amplified in self-refinement pipelines, which enhance fluency and coherence but reinforce the model's preference for its own outputs. Building on these findings, \citet{Panickssery2024} investigate the underlying mechanisms of LLM self-preferencing and find that a model's ability to recognize its own outputs, i.e., its self-recognition capability, contributed significantly to this bias. Specifically, LLMs such as GPT-4 and LLaMA 2 exhibite non-trivial self-recognition capabilities, and there is a strong positive correlation between a model's self-recognition capability and the magnitude of its self-preference bias.

While prior work has established the existence of self-preference bias in LLM benchmarks, its broader consequences in business operations, particularly in high-stakes scenarios like algorithmic hiring, remain underexplored. This study builds on existing research by empirically measuring self-preference bias in algorithmic hiring processes. By extending the discussion into real-world decision-making contexts, this study provides critical insights for AI governance, ethical AI deployment, and the design of unbiased algorithmic hiring frameworks.



\section{Definition and Measurement of AI Self-Preference Bias}
\label{section: methodology}
In this section, we formally define AI self-preference bias and present empirical strategies to quantify the extent of this bias in the context of algorithmic hiring.  

%
\subsection{Definition of AI Self-Preference Bias}\label{sec:def_self_pref_bias}
Building on recent work \citep{Panickssery2024, Wataoka2024}, we conceptualize AI self-preference bias as the tendency of an LLM to favor content it generated itself over content from other sources. This bias can manifest in two distinct forms:
\begin{enumerate}
    \item LLM-vs-Human Self-Preference: The tendency of an LLM to prefer its own generated content over human-written content.
    \item LLM-vs-LLM Self-Preference: The tendency of an LLM to prefer its own output over content generated by an alternative LLM.
\end{enumerate}

\begin{figure}[t!]
\centering
\begin{subfigure}[t]{0.5\linewidth}
    \centering
    \includegraphics[width=\linewidth]{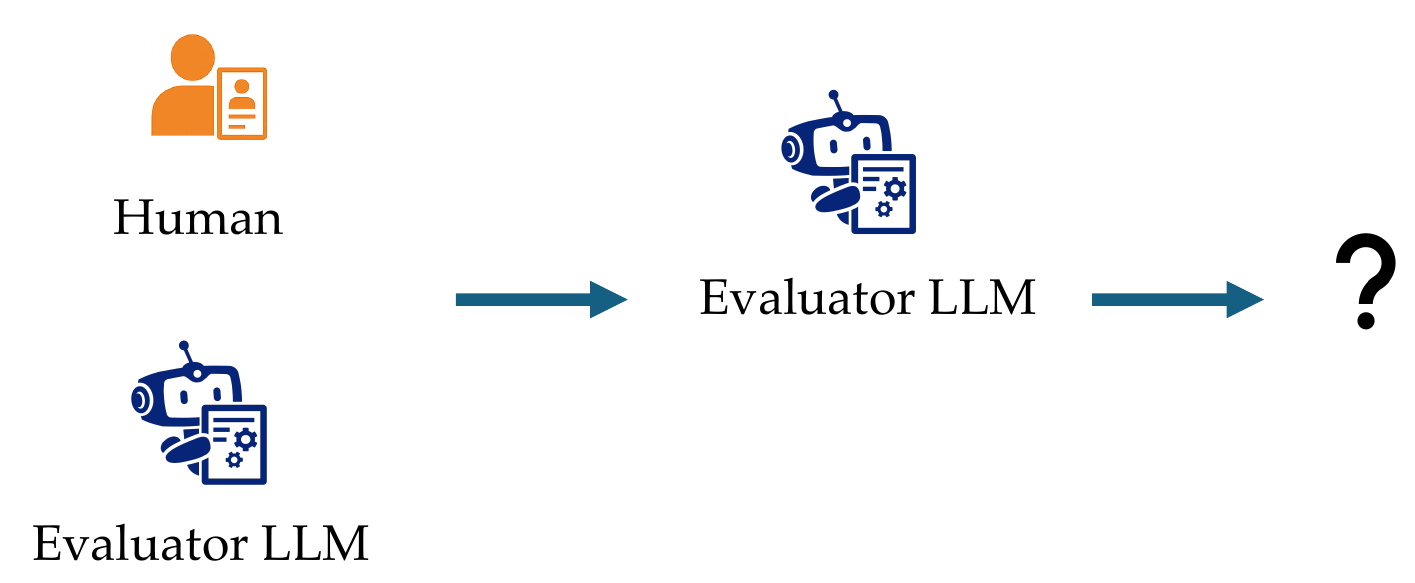}
    \caption{LLM-vs-Human Self-preference Bias}
    \label{fig:LLM-Human-bias}
\end{subfigure}
\vfill
\begin{subfigure}[t]{0.5\linewidth}
    \centering
    \includegraphics[width=\linewidth]{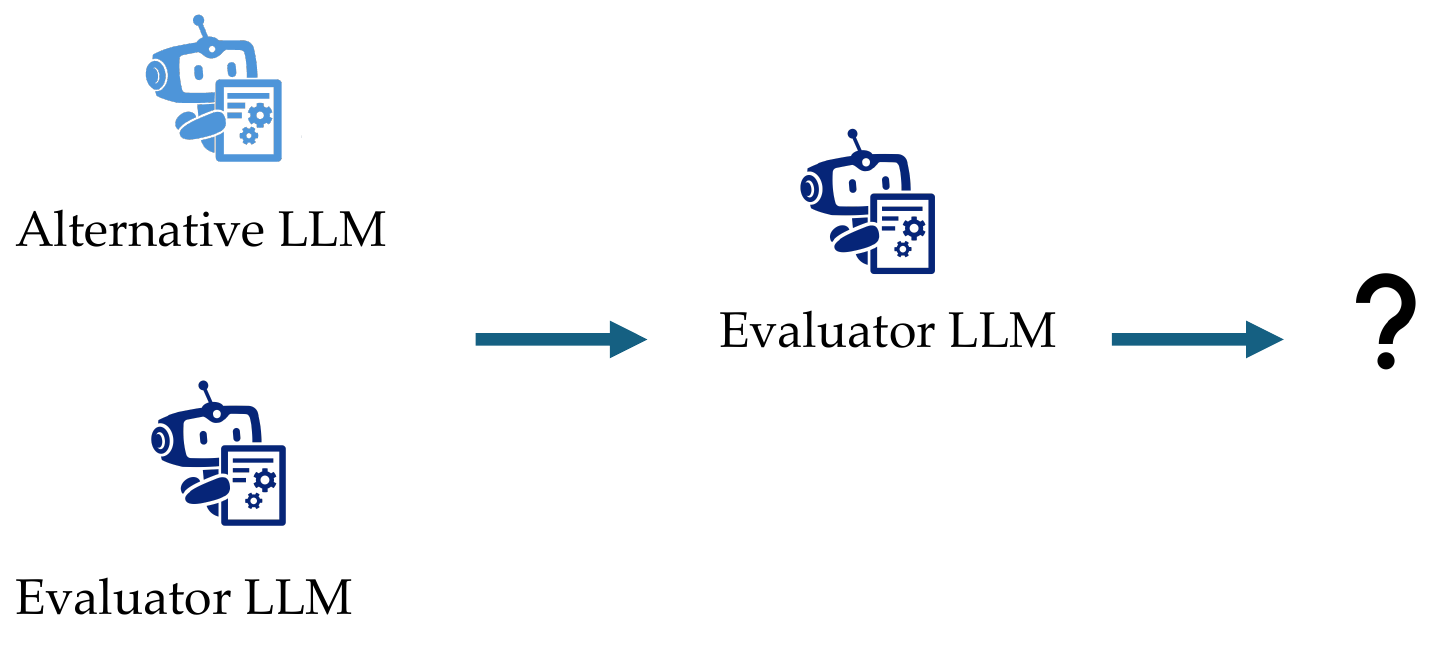}
    \caption{LLM-vs-LLM Self-preference Bias}
    \label{fig:LLM-LLM-bias}
\end{subfigure}
\caption{Illustration of the context of AI self-preference bias. In both cases, the evaluator LLM makes a choice between paired resumes, allowing us to test whether it systematically favors its own outputs.}
\label{fig:self-bias}
\end{figure}



To empirically evaluate self-preference bias, we test whether a given LLM, acting as the evaluator (the evaluator LLM), makes unbiased selections when presented with carefully matched \textit{pairs} of resumes (\autoref{fig:self-bias}). Each pair represents the \emph{same underlying candidate}, with identical qualifications, experience, and background information, differing only in how that information is expressed. In the LLM-vs-human case (\autoref{fig:LLM-Human-bias}), the evaluator LLM compares a human-written resume to a counterfactual version it generated itself, with both describing the same candidate profile. In the LLM-vs-LLM case (\autoref{fig:LLM-LLM-bias}), the evaluator LLM compares its own generated version to one produced by an alternative LLM. Under this design, the evaluator LLM's task reduces to choosing between two alternative representations of the same candidate.




\subsection{Quantifying AI Self-Preference Bias}
When deployed in algorithmic hiring contexts, an LLM performing such pairwise comparisons effectively functions as a binary classifier. The measurement of bias in such classifiers has been extensively studied in the algorithmic fairness literature (e.g., \citealt{calders2009building, hardt2016equality}). Building on this framework, we analyze both forms of self-preference through the lens of two foundational fairness criteria: \textit{statistical parity} \citep{calders2009building} and \textit{equal opportunity} \citep{hardt2016equality}, which capture distinct notions of fairness in algorithmic decision-making.
\subsubsection{Statistical Parity} ~\

Statistical parity requires that the probability of a positive outcome---in this context, the likelihood of a resume being selected as better---be equal across groups defined by a focal attribute. While this criterion is traditionally applied to protected human attributes such as gender or race, we adapt it here to study disparities across resumes defined by their source of generation: whether a resume was generated by the evaluator LLM or by an alternative source (a human or another LLM). 
In this setting, statistical parity captures unconditional differences in selection rates and serves as a descriptive measure of whether the evaluator LLM disproportionately selects its own outputs.

To operationalize this, we conduct pairwise resume comparisons involving resumes generated by either a human or one of several LLMs. Specifically, we test whether an LLM $f$, when serving as the evaluator LLM, is more likely to select a resume it generated over one written by a human, or over one produced by an alternative LLM, when the resumes are otherwise equivalent in content quality. 

Let $S \in \{0, 1\}$ denote the source indicator, where $S = 1$ if the resume was generated by the evaluator LLM $f$, and $S = 0$ if it was written by a human or generated by an alternative LLM. Let $Y_f^{\prime} \in \{0, 1\}$ denote the binary decision made by evaluator $f$, where $Y_f^{\prime} = 1$ indicates that the resume is selected as stronger. We define the Statistical Parity Self-Preference Bias as the difference in \textit{selection rates} between resumes generated by the evaluator LLM and those from other sources:
\begin{align}
\label{eq:AI_self_bias}
\text{Statistical Parity Self-Preference Bias}_f &= \mathbb{P}\left(Y_f^{\prime} = 1 \mid S = 1\right) - \mathbb{P}\left(Y_f^{\prime} = 1 \mid S = 0\right).
\end{align}

This formulation Eq.~(\ref{eq:AI_self_bias}) captures the difference in the evaluator LLM $f$'s likelihood of selecting a resume it generated versus one it did not, providing a direct measure of unconditional AI self-preferencing. Interpreting this difference as bias, however, requires caution because it can reflect not only self-preference but also systematic differences in \emph{content quality}. In this context, content quality refers to the effectiveness with which a resume communicates the same underlying candidate information---such as clarity, coherence, fluency, and organization---rather than differences in the candidate's qualifications or background. As a result, a positive statistical-parity gap may arise even in the absence of intrinsic self-preference if the evaluator LLM tends to produce higher-quality renderings of the same candidate information (for instance, when users rely on a more capable model to polish resume wording and structure).

\subsubsection{Equal Opportunity} ~\

To address the limitation of the statistical parity measure and disentangle intrinsic self-preferencing from differences in content quality, we adopt the equal opportunity fairness criterion of \citet{hardt2016equality}. The key idea underlying equal opportunity is to compare selection behavior conditional on merit. In our setting, this allows us to assess whether an evaluator LLM exhibits systematic self-preference even when resumes are of comparable content quality.

Let $Y \in \{0, 1\}$ represent the ground truth quality label, where $Y = 1$ indicates that the resume is of higher content quality and would be selected by a human evaluator. We define the Equal Opportunity Self-Preference Bias for evaluator LLM \(f\) as:
\begin{align}
\label{eq:AI_self_bias_quality_adjusted}
\text{Equal Opportunity Self-Preference Bias}_f &= \mathbb{P}\left(Y_f^{\prime} = 1 \mid S = 1, Y = 1\right) - \mathbb{P}\left(Y_f^{\prime} = 1 \mid S = 0, Y = 1\right).
\end{align}
By conditioning on resume content quality, this formulation Eq.~(\ref{eq:AI_self_bias_quality_adjusted}) isolates the evaluator LLM $f$'s intrinsic tendency to favor its own generated content. A positive value of this measure indicates that, even among resumes that are comparably high quality, the evaluator LLM $f$ is more likely to select resumes it generated itself, providing evidence of self-preferencing that cannot be explained by superior content quality alone.


We operationalize the equal opportunity criterion using two complementary approaches that reflect the fact that content quality is a latent variable and cannot be directly observed. First, we use a regression model to control for observable linguistic and textual features, which serve as proxies for content quality in pairwise comparisons. Second, we rely on human annotators to independently evaluate each resume pair and determine which resume is better, using these judgments as the ground truth for the latent content quality.  

\textbf{Conditional Logistic Regression.} In line with the rationale behind the equal opportunity, we estimate a conditional logistic regression model that conditions on each resume pair $\mathcal{R}_i$, thereby absorbing all pair-specific quality differences. Let $i$ index resume pairs and $j \in \{1,2\}$ index the two resumes within each pair. The model is given by
\begin{align}
\label{eq:logistic_reg}
\log\left(\dfrac{\mathbb{P}(\text{Preferred}_{ij} = 1|\mathcal{R}_i)}{1 - \mathbb{P}(\text{Preferred}_{ij} = 1|\mathcal{R}_i)}\right) =\beta_1 \cdot \text{evaluatorLLM}_{ij}  + \underbrace{\boldsymbol{\beta}_2^{\top} \boldsymbol{\phi}_{ij} + \boldsymbol{\beta}_3^{\top}\boldsymbol{\psi}_{ij}}_\textrm{Content Quality Controls},
\end{align}
where $\text{Preferred}_{ij}=1$ indicates that resume $j$ is selected as better within pair $i$, and $\text{evaluatorLLM}_{ij}$ is an indicator for whether resume $j$ was generated by the evaluator LLM. Standard errors are clustered at the resume-pair level to account for within-pair correlation in residuals.


We include two sets of control variables that proxy for the latent quality of resume content. The first set, denoted by $\boldsymbol{\phi}_{ij}$, comprises a rich collection of linguistic features drawn from the Linguistic Inquiry and Word Count (LIWC) framework \citep{boyd2022development}. These features capture multiple dimensions of writing style and psychological tone that are plausibly related to how effectively a resume conveys information. Specifically, we include measures of (i) length and lexical complexity, such as total word count and the prevalence of longer words; (ii) psychological tone and professionalism, captured by Analytic, Clout, Authentic, and overall emotional Tone; (iii) linguistic structure, reflected in the use of function words and core grammatical categories (e.g., prepositions, auxiliary verbs, and conjunctions); and (iv) agency and action orientation, proxied by the use of personal pronouns (e.g., I, we, you, they) and the balance between verbs and adjectives.



The second set of controls, denoted by $\boldsymbol{\psi}_{ij}$, captures semantic similarity and content preservation between the resume summary and the rest of the resume. We include BERTScore \citep{Tianyi2020}, which measures semantic similarity using contextual embeddings from BERT, and ROUGE-L \citep{lin-2004-rouge}, which captures sequence-level overlap based on longest common subsequences. These metrics help ensure that estimated effects are not driven by differences in informational content or semantic drift, but rather reflect differences in presentation and stylistic expression.

In the specification given in Eq. (\ref{eq:logistic_reg}), the key parameter of interest is $\beta_1$, which captures the evaluator LLM's tendency to prefer its own output, after controlling for observable content quality. A significantly positive estimate of $\beta_1$ indicates that the evaluator LLM is more likely to selects its own output over competing content, even after adjusting for measurable aspects of content quality. With this parameter, the equal opportunity self-preference bias can be computed with $\frac{e^{\beta_1}}{1+e^{\beta_1}}-\frac{1}{1+e^{\beta_1}}$, which represents the difference in the probability that an LLM evaluator prefers its own output over a competing alternative, holding content quality constant. 


\textbf{Human Annotations.} Since measuring equal-opportunity self-preference bias requires ground-truth labels identifying which resume in a pair is of higher quality, we obtain human evaluations.

To this end, we recruit $18$ human annotators from Prolific to evaluate two types of resume pairs: (1) human-written resumes versus their LLM-generated counterfactuals, and (2) resumes generated by one LLM versus another LLM. For both tasks, we focus on three representative LLMs: GPT-4o, DeepSeek-V3, and LLaMA-3-70B. Specifically, for case (1), the evaluation includes (i) GPT-4o vs. Human, (ii) DeepSeek-V3 vs. Human, and (iii) LLaMA-3-70B vs. Human. For case (2), we evaluate (i) DeepSeek-V3 vs. GPT-4o, (ii) DeepSeek-V3 vs. LLaMA-3-70B, and (iii) GPT-4o vs. LLaMA-3-70B.
Each comparison condition is evaluated by three annotators.

Each annotator is presented with $30$ resume pairs from one of the comparison cases above.\footnote{In the annotation instructions, annotators were informed that there are 32 pairs in total, including 2 attention-check pairs that are excluded from the final analysis.} The number of evaluations per annotator is chosen to balance cognitive workload with fair compensation. 
To prevent confirmation bias, the order of the resumes within each pair is randomized, and annotators are blinded to the source of each resume. For each resume pair, annotators rate both resumes on five linguistic dimensions---clarity, fluency, coherence, conciseness, and overall quality---and are then asked to select which resume is better. 

After obtaining the quality label from human annotators, we apply bootstrapping with $10,000$ resamples to estimate the equal opportunity self-preference bias. 
The annotation interface and instructions can be found in \aref{appendix:interface} and \aref{appendix:instructions}. 

\section{Data and Experimental Design}
\label{section: data}
To empirically examine AI self-preference bias as defined in \autoref{section: methodology}, we design a series of resume correspondence experiments \citep{bertrand2004emily} in which LLMs act as evaluators (evaluator LLM) tasked with screening and selecting between pairs of resumes. In each pair, one resume is generated by the evaluator LLM itself, while the other is written by a human or produced by an alternative LLM. This setup allows us to test whether an evaluator LLM systematically favors its own outputs over others.

To construct these resume pairs, we begin with a dataset of human-written resumes. For each human-written resume, we use several state-of-the-art LLMs to generate counterfactual versions that describe the same candidate profile. We then form multiple types of resume pairs: evaluator LLM-generated versus human-written, and evaluator LLM-generated versus alternative LLM-generated. These pairs serve as the input for the evaluator LLMs.

In the subsections that follow, we first describe the original resume dataset. We then outline our experimental procedures, including the generation of counterfactual resumes and the design of the pairwise comparisons.

\subsection{Data}
We draw on a publicly available dataset containing $2{,}484$ anonymized, human-written resumes collected from the professional resume-building platform LiveCareer.com \citep{bhawal2021resume}. The resumes were authored by real job seekers prior to the widespread adoption of large language models, ensuring that the content reflects human-written resumes rather than AI-generated text and thus making it well-suited for our study of AI self-preferencing. The dataset has been widely used in recent research on AI and algorithmic hiring. For example, it has served as a benchmark for evaluating hiring biases in language models \citep{Wang-etal-2024-jobfair}, and has been used to investigate gender, racial, and intersectional biases in resume screening using language model retrieval techniques \citep{Wilson_Caliskan_2024}.

Spanning $24$ distinct occupational categories, including teachers, consultants, chefs, engineers, and more, the dataset offers a diverse representation of professional backgrounds. Each resume typically includes multiple sections: an executive summary, education, work experience, and skills. Among these, the executive summary is particularly relevant to our study. This section comprises a free-text narrative in which candidates synthesize their qualifications, accomplishments, and career objectives, and is often used by evaluators to form initial impressions. Unlike structured sections such as education or work experience, which largely report verifiable facts, the executive summary is inherently subjective and stylistically flexible. As a result, it offers substantial scope for variation in phrasing, tone, and presentation---dimensions along which LLMs may exert influence. This makes the executive summary a natural and empirically relevant focus for examining how AI-generated text shapes perceptions and potentially affects hiring outcomes.

Thus, we focus our analysis on the executive summary section of each resume. To systematically construct counterfactual resumes, we replace the original executive summary of each human-written resume with an LLM-generated version, while preserving all other content (e.g., work experience, skills, education) unchanged. This approach isolates the effect of LLM-generated text on candidate evaluation outcomes, holding constant the factual and objective components of the resume. By doing so, we avoid potential confounds introduced by allowing LLM to modify structured information, which could lead to hallucinations or factual inaccuracies.

\begin{table}[tb]
\centering
\caption{Summary Statistics of Resume Summaries}
\label{tab:stacked_summary_stats}
\resizebox{\textwidth}{!}{\begin{tabular}{lccccccc}
\toprule
Measure & Mean & Std. Dev. & Min & 1st Quartile & Median & 3rd Quartile & Max \\
\midrule
Number of words & 70.74 & 72.05 & 3.00 & 32.00 & 50.00 & 84.00 & 1216.00 \\
Number of sentences & 3.85 & 4.11 & 1.00 & 2.00 & 3.00 & 5.00 & 61.00 \\
Average words per sentence & 21.78 & 17.33 & 1.50 & 14.00 & 18.00 & 24.00 & 369.00 \\
Number of unique words & 52.53 & 39.32 & 2.00 & 28.00 & 42.00 & 64.00 & 520.00 \\
Type-Token Ratio & 0.82 & 0.10 & 0.32 & 0.75 & 0.83 & 0.89 & 1.00 \\
Presence of numbers & 0.37 & 0.48 & 0.00 & 0.00 & 0.00 & 1.00 & 1.00 \\
\bottomrule
\end{tabular}}
\begin{minipage}{\linewidth}
\vspace{1ex}
\footnotesize \textit{Notes:} The Type-Token Ratio (TTR) is a commonly used measure of lexical diversity, calculated as the number of unique words (types) divided by the total number of words (tokens) in a text. 
\end{minipage}
\end{table}

This dataset forms the basis of our experimental design. To construct a clean and comparable set of resumes, we preprocess the data by extracting the executive summary sections, removing formatting artifacts, and excluding observations with empty or missing summaries. After cleaning and validation, the final sample consists of $2{,}245$ resumes, which are used in our correspondence experiments. \autoref{tab:stacked_summary_stats} reports descriptive statistics for the human-written executive summaries. These summaries exhibit substantial variation in word and sentence counts, reflecting heterogeneity in writing styles across resumes. Measures of lexical diversity, including the type–token ratio, further indicate meaningful variation in vocabulary usage across summaries. \footnote{While there is no universally accepted threshold for high lexical diversity or an optimal type-token ratio, values above $0.5$ are often considered indicative of high lexical diversity in short texts.
} 

\subsection{Experimental Design}
To examine AI self-preference bias in the context of algorithmic hiring, we design a series of resume correspondence experiments comprising two steps: counterfactual resume generation and pairwise resume evaluation, as illustrated in the \autoref{fig:experiment_design}.


\subsubsection{Counterfactual Resume Generation}~\


We evaluate a diverse set of large language models that vary along two dimensions relevant to real-world hiring and resume-writing contexts: access regime (closed- versus open-source) and model scale. Specifically, we consider three closed-source models (GPT-4o, GPT-4o-mini, and GPT-4-turbo) and four open-source models (LLaMA~3.3-70B,\footnote{The suffix (e.g., ``70B'') refers to the number of parameters in the model---in this case, 70 billion. Larger models generally have greater capacity for language understanding and generation, though size alone does not determine overall performance.} Mistral-7B, Qwen~2.5-72B, and DeepSeek-V3). These models represent widely used and competitive systems for text generation and summarization tasks and exhibit broadly comparable performance on standard summarization benchmarks, as reported by the ProLLM leaderboard \citep{prollm2024leaderboard}. Our analysis does not rely on exact parity in benchmark performance; rather, these models capture a relevant cross-section of contemporary LLMs that job seekers or platforms may plausibly employ in resume preparation. To examine how model scale relates to self-preferencing behavior, we additionally include two smaller-scale language models---LLaMA~3.2-1B and LLaMA~3.2-3B. These models are not intended to represent tools commonly used in hiring contexts, but instead allow us to explore whether self-preference varies systematically with model capacity.

To generate resume summaries, we prompt each LLM with a modified version of the original resume in which the human-written summary is removed, but all other sections (e.g., work experience, skills, education) are left unchanged. The prompted LLM then generates a new resume summary, which is inserted back into the original resume to form a complete counterfactual version. By construction, each counterfactual resume differs from the original only in the linguistic expression of the executive summary, holding constant all underlying candidate attributes. The full prompt used for summary generation is provided in \aref{appendix:prompts}.

To control for verbosity bias \citep{saito2023verbosity}, where LLM evaluators may favor longer, verbose responses, we explicitly instruct each LLM to generate summaries within a specified word range, corresponding to the $1^{st}$ and $3^{rd}$ quartiles of the empirical length distribution of human-written summaries. This constraint minimizes variation in length across models and prevents length from confounding downstream evaluations. 

\begin{figure}[htbp]
    \centering
    \begin{subfigure}[b]{0.7\linewidth}
        \centering
        \includegraphics[width=\linewidth]{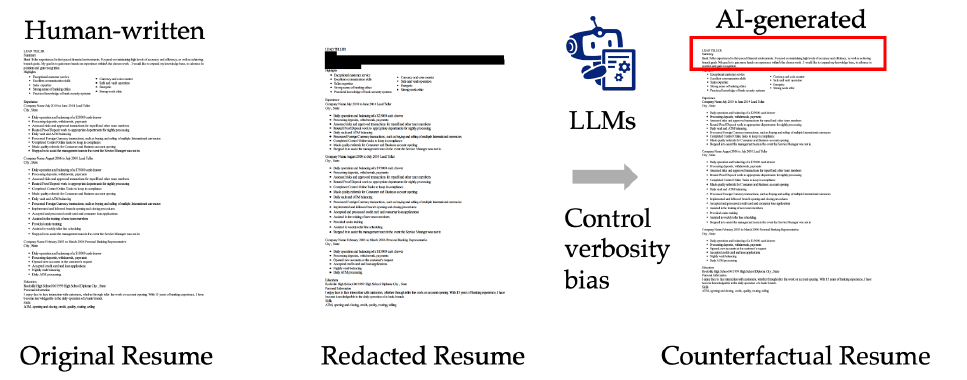}
         \caption{Counterfactual Resume Generation}
         \label{fig:resume-generation}
    \end{subfigure}
    \vfill
    \begin{subfigure}[b]{0.7\linewidth}
        \centering
        \includegraphics[width=\linewidth]{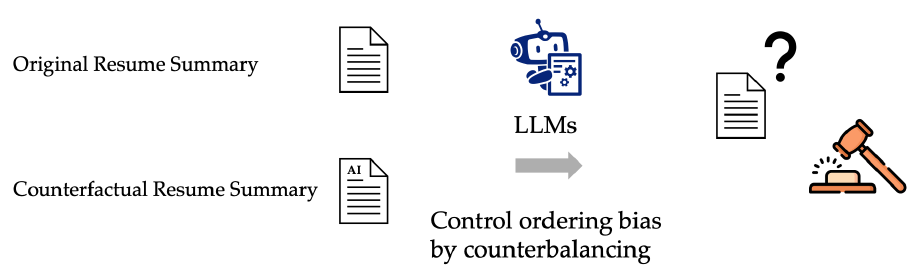}
           \caption{Pairwise Resume Evaluation}
        \label{fig:resume-evaluation}
    \end{subfigure}
    \caption{Resume Correspondence Experiments}
    \label{fig:experiment_design}
\end{figure}

The generation process described above is illustrated in \autoref{fig:resume-generation}. For each original human-written resume, each LLM produces a corresponding counterfactual version that differs only in the executive summary section. All other resume components are held fixed. This design isolates the effect of the \emph{source of the summary text}---whether it is human-written, generated by the evaluator LLM, or generated by an alternative LLM.  In the next subsection, we describe how these counterfactual resumes are paired and evaluated to systematically test for AI self-preference bias.

\subsubsection{Pairwise Resume Evaluation} ~\

To test for AI self-preference bias, we conduct pairwise evaluations based on the executive summary text. Although counterfactual summaries are generated conditional on the full resume to ensure that all versions are grounded in the same underlying candidate information, evaluator LLMs are shown only the two executive summaries when making their selection. Since all other resume content is identical across counterfactual versions, it is omitted from the evaluation stage.

Focusing the evaluation on the executive summary ensures that comparisons are made between alternative textual representations of the same candidate profile, without introducing additional variation from other resume sections. This design allows us to attribute differences in evaluation outcomes to the source and expression of the summary text, rather than to differences in underlying candidate information or document-level alignment.

To simulate a resume screening task, we designate one LLM as the evaluator LLM and refer to all other models as alternative LLMs. As illustrated in \autoref{fig:resume-evaluation}, the evaluator LLM is prompted to compare two executive summaries and select the stronger one based on the overall presentation of skills and experience, following a standardized, job-agnostic evaluation instruction. For each evaluator LLM, we construct two types of summary pairs: (1) a summary generated by the evaluator LLM versus a human-written summary, and (2) a summary generated by the evaluator LLM versus one generated by an alternative LLM. The full prompt for the evaluation task is provided in Appendix~\ref{appendix:prompts}.

To account for position (or ordering) bias \citep{zheng2023judging}, where LLMs may exhibit a preference for the first or second option presented, we implement a counterbalanced design following \cite{brooks2012counterbalancing}. Specifically, the order of resumes (i.e., A vs. B) is randomized across comparisons. This approach ensures that any position-related biases are evenly distributed and do not systematically affect evaluation outcomes, thereby preserving internal validity without requiring duplicated evaluations. See \aref{appendix:summary_example} for examples of both human-written and LLM-generated professional summaries used in the pairwise comparisons.

\section{Empirical Results}
\label{section: empirical_results}
Building on the definition and measurements of AI self-preference bias introduced in \autoref{section: methodology}, we empirically evaluate the extent to which LLMs favor their own generated content over that produced by humans or alternative LLMs. We begin by analyzing the LLM-vs-Human self-preference, followed by the LLM-vs-LLM self-preference. For each form of bias, we present results based on the two fairness metrics (statistical parity and equal opportunity) to provide complementary perspectives on self-preferencing behavior. 

\subsection{LLM-vs-Human Self-Preference}
We find strong and consistent evidence that LLMs, when used as evaluators, systematically prefer resumes they generated themselves over equivalent resumes written by humans. This bias is pervasive across models and persists even after controlling for content quality. Notably, the strength of self-preference increases with model size, which may indicate that larger models are more sensitive to stylistic features resembling their own outputs.

\subsubsection{Statistical Parity Self-Preference Bias} ~\

\begin{figure}[t!]
\centering
\begin{subfigure}[t]{0.8\linewidth}
    \centering
    \includegraphics[width=\linewidth]{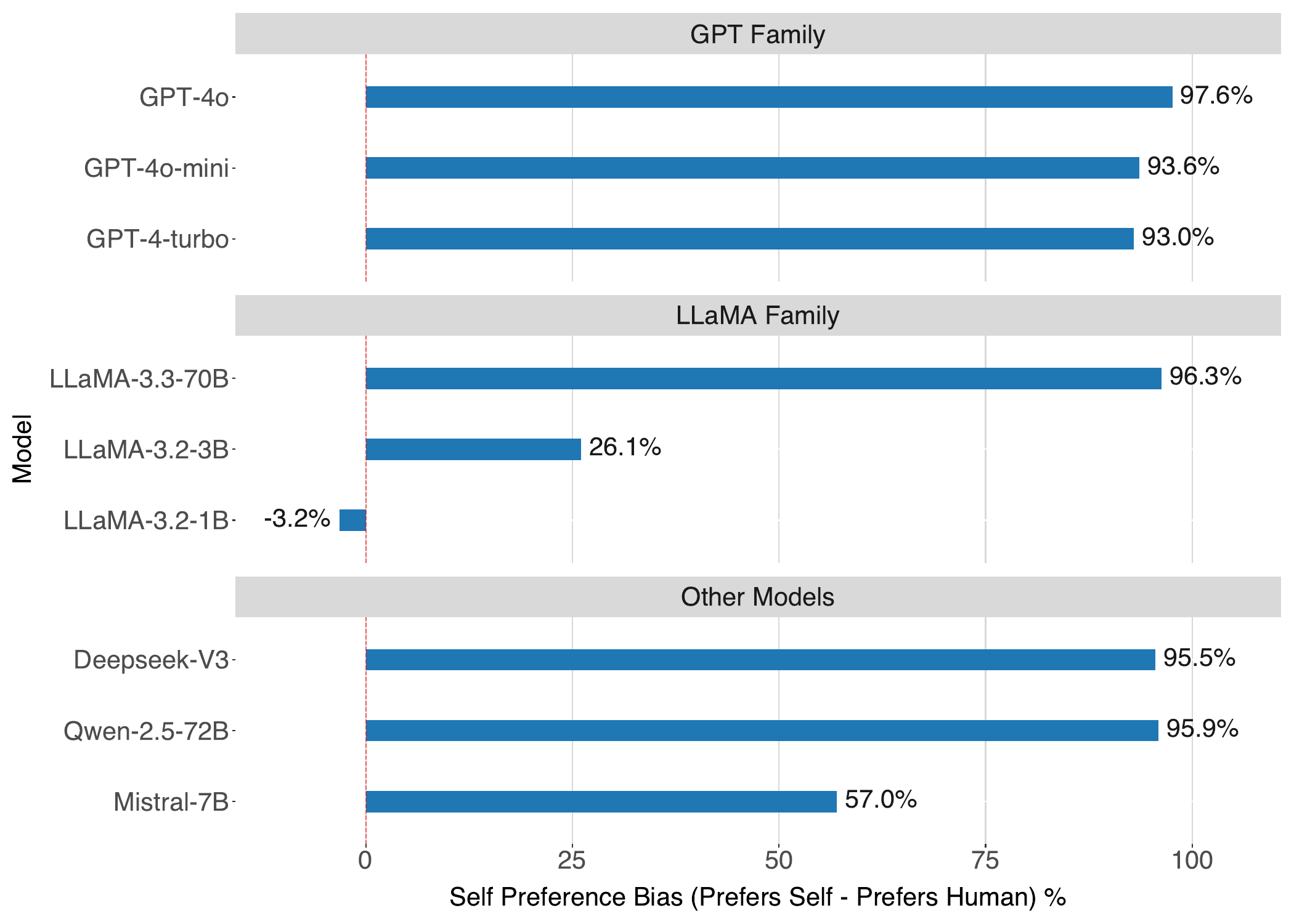}
    \caption{LLM-vs-Human Statistical Parity Self-preference Bias}
    \label{fig:ai-self-bias}
\end{subfigure}
\vfill
\begin{subfigure}[t]{0.8\linewidth}
    \centering
    \includegraphics[width=\linewidth]{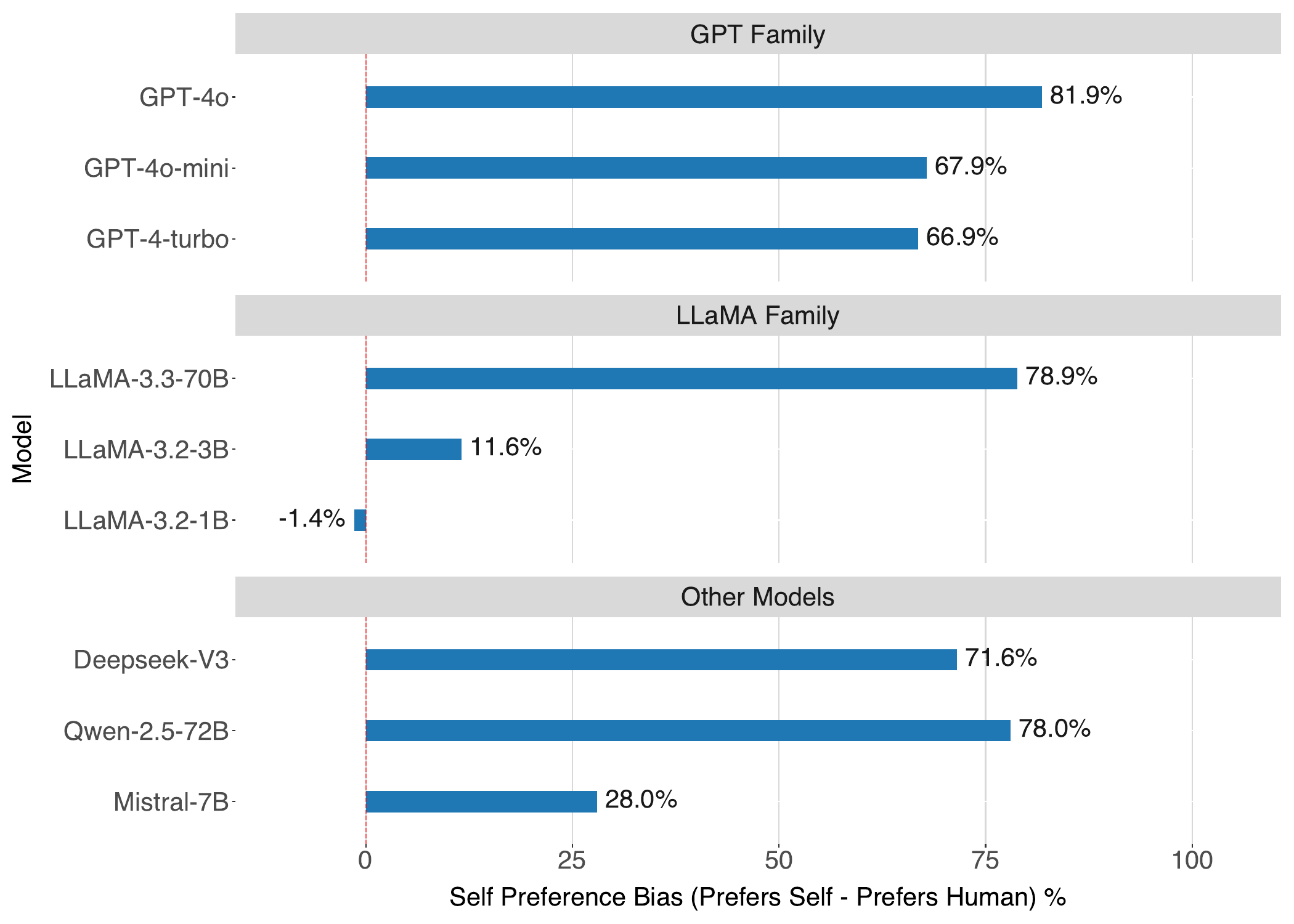}
    \caption{LLM-vs-Human Equal Opportunity Self-preference Bias}
    \label{fig:ai-self-EO-bias}
\end{subfigure}
\caption{LLM-vs-Human Self-Preference Bias}
\label{fig:self-bias-combined}
\end{figure}

Under the statistical parity metric defined in Eq.~(\ref{eq:AI_self_bias}), we compare the probability that an evaluator LLM selects its own generated resume with the probability that it selects a human-written resume. The difference between these probabilities provides a direct measure of self-preference bias: positive values indicate systematic favoritism toward the model's own outputs, values near zero reflect parity, and negative values indicate a preference for human-written resumes. As shown in \autoref{fig:ai-self-bias}, each bar reports the outcome of an evaluator LLM choosing between a human-written resume and a version it generated itself. Eight of the nine LLMs tested exhibit clear LLM-vs-Human self-preference, with magnitudes ranging from $26\%$ to $98\%$. In practical terms, this means evaluator LLMs are between $26\%$ and $98\%$ more likely to select a resume they generated themselves than an equivalent resume written by a human.

\subsubsection{Equal Opportunity Self-Preference Bias} ~\

To evaluate equal opportunity, we assess whether self-preference persists when resumes are of equivalent quality, as defined in Eq.~(\ref{eq:AI_self_bias_quality_adjusted}).  We use two complementary approaches: (1) a conditional logistic regression controlling for content quality features, and (2) human annotations indicating which resume is objectively better. 

As shown in \autoref{fig:ai-self-EO-bias}, after adjusting for content quality using conditional logistic regression, most evaluator models continue to exhibit significant self-preferencing behavior. Consistent with the statistical parity results, all models except LLaMA-3.2-1B display positive and significant self-preference bias. Larger models---such as GPT-4o, GPT-4-turbo, DeepSeek-V3, Qwen-2.5-72B, and LLaMA-3.3-70B---exhibit the strongest LLM-vs-Human self-preference, with magnitudes ranging from $67\%$ to $82\%$. In contrast, smaller models display weaker biases: Mistral-7B and LLaMA-3.2-3B show modest but statistically significant effects ($28\%$ and $11.6\%$, respectively). More detailed conditional logistic regression results from Eq.~(\ref{eq:logistic_reg}) are presented in \aref{appendix:regression_results} (\autoref{tab:logistic_reg_1st_bias}).  

We further assess equal opportunity using human-annotated ground truth. 
Strikingly, across all three evaluator LLMs with human-annotated comparisons (GPT-4o, DeepSeek-V3, and LLaMA 3.3-70B), each model consistently selected its own generated summary over the human-written alternative—even in cases where human annotators judged the human-written summary to be higher quality in its presentation of the same underlying candidate  (e.g., clearer, more coherent, or more effective overall).

\subsection{LLM-vs-LLM Self-Preference}
We next examine whether self-preference extends to cases where LLMs evaluate content generated by alternative LLMs. Specifically, we focus on GPT-4o, DeepSeek-V3, and LLaMA 3.3-70B, three of the most advanced and widely adopted models in both commercial and research contexts. These models not only achieve state-of-the-art performance across benchmarks \citep{hurst2024gpt, deepseekai2024deepseekv3technicalreport, dubey2024llama} but also represent diverse development lineages (OpenAI, DeepSeek, and Meta, respectively), making them especially relevant for understanding how self-preference may manifest across different model families and design philosophies.  In this setting, each model serves as the evaluator LLM and is presented with pairwise comparisons between two counterfactual resumes: one generated by itself and the other by a competing LLM. Both resumes are derived from the same underlying human-written original, ensuring comparability of content. In contrast to the consistent self-preferencing bias documented in LLM-vs-Human comparisons, LLM-vs-LLM self-preferencing behavior proves considerably more heterogeneous across models.

\subsubsection{Statistical Parity Self-Preference Bias} ~\

\begin{figure}[t!]
\centering

\begin{subfigure}[t]{\textwidth}
  \centering
 \includegraphics[width=0.32\linewidth]{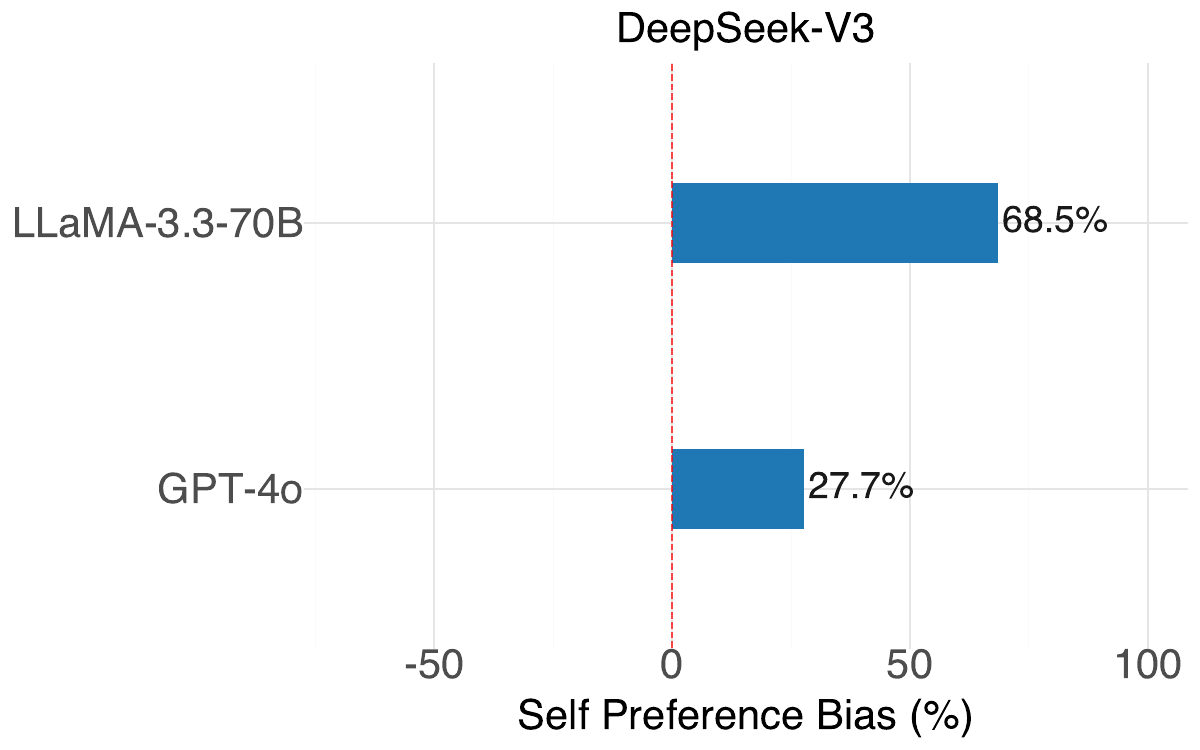}\hfill
  \includegraphics[width=0.32\linewidth]{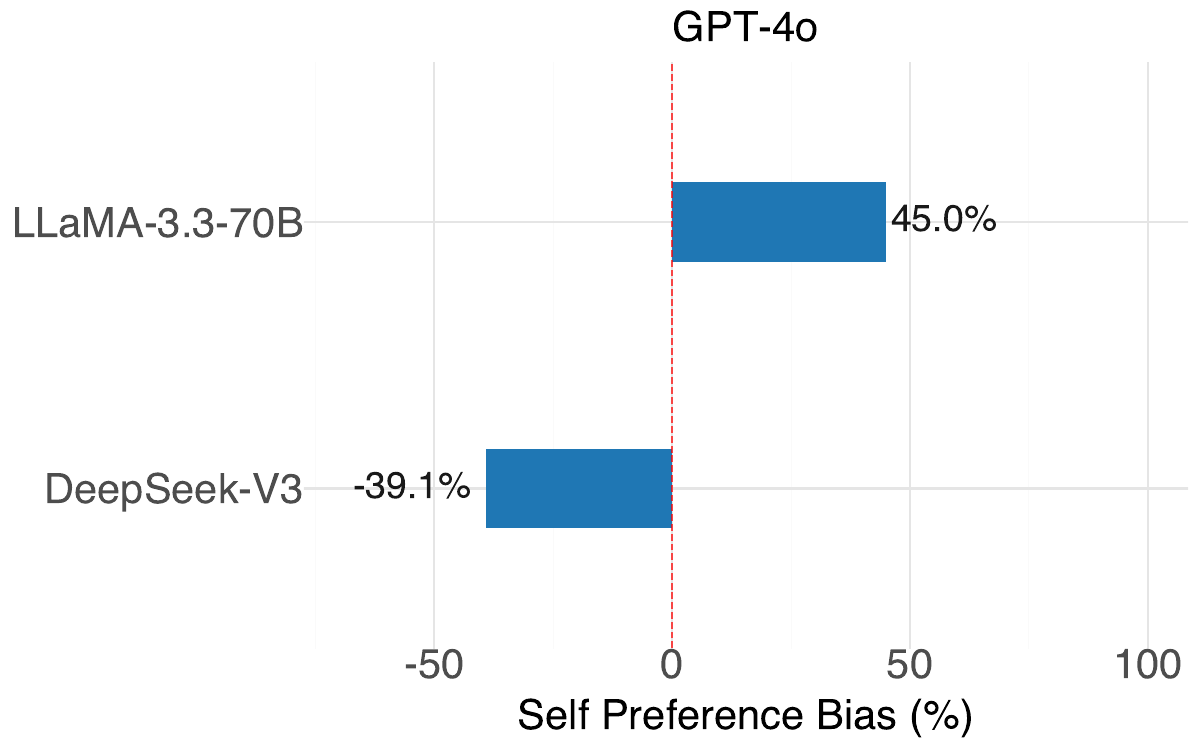}\hfill
  \includegraphics[width=0.32\linewidth]{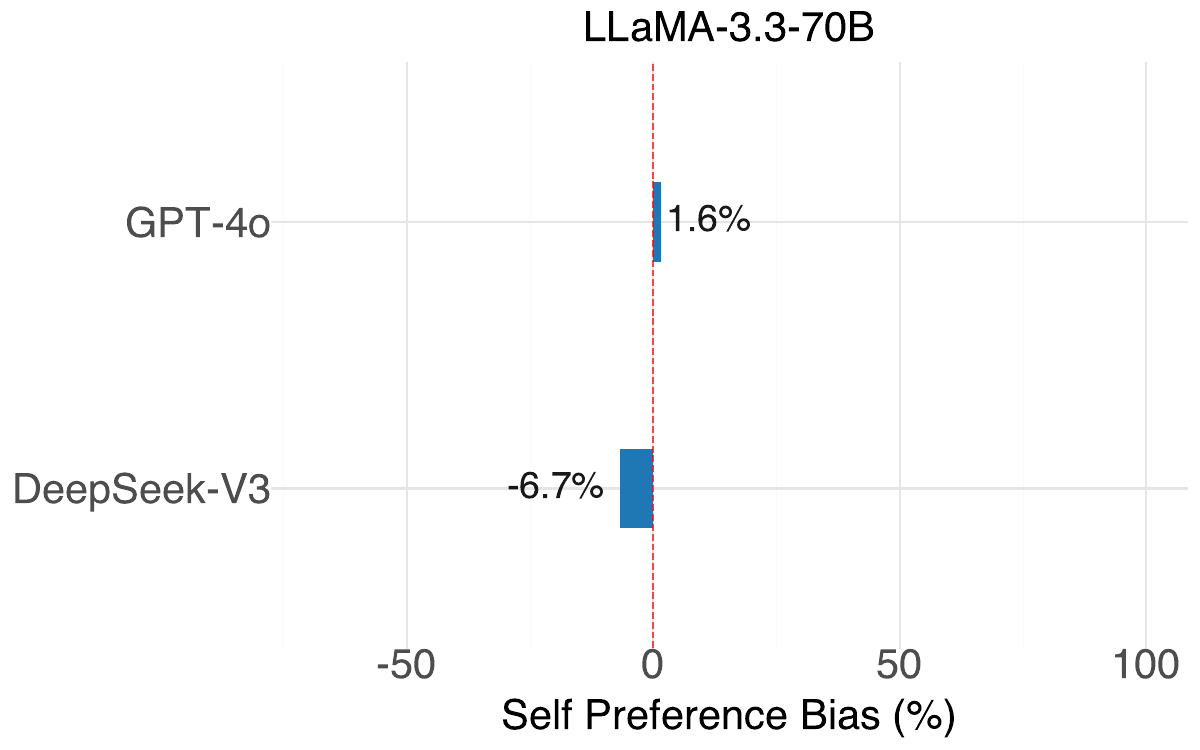}
  \caption{LLM-vs-LLM Statistical Parity Self-preference Bias}
  \label{fig:sp-row}
\end{subfigure}

\vspace{2ex}

\begin{subfigure}[t]{\textwidth}
  \centering
  \includegraphics[width=0.32\linewidth]{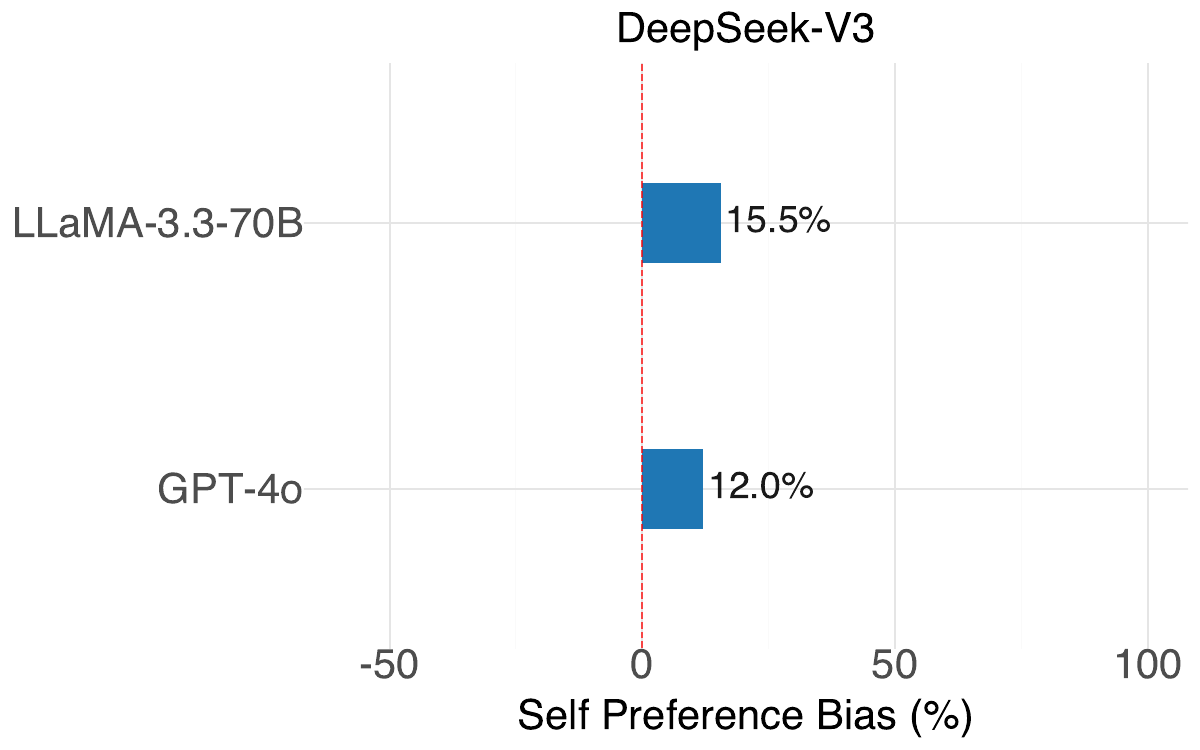}\hfill
  \includegraphics[width=0.32\linewidth]{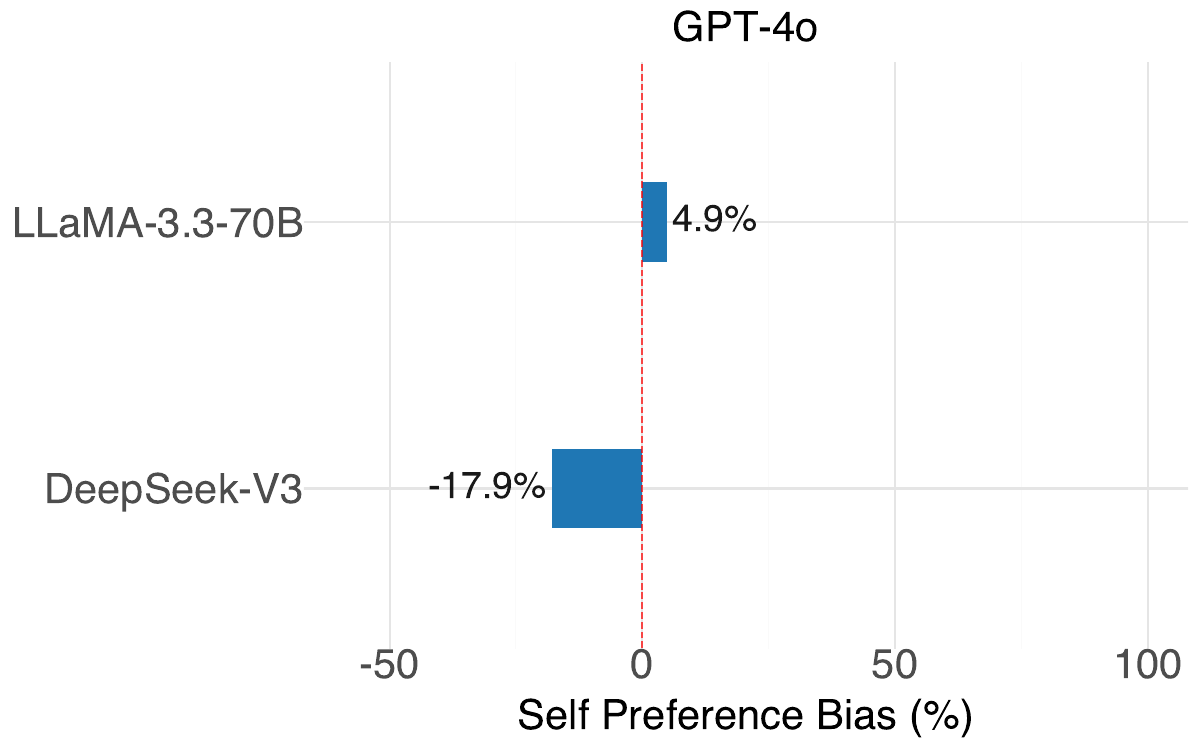}\hfill
  \includegraphics[width=0.32\linewidth]{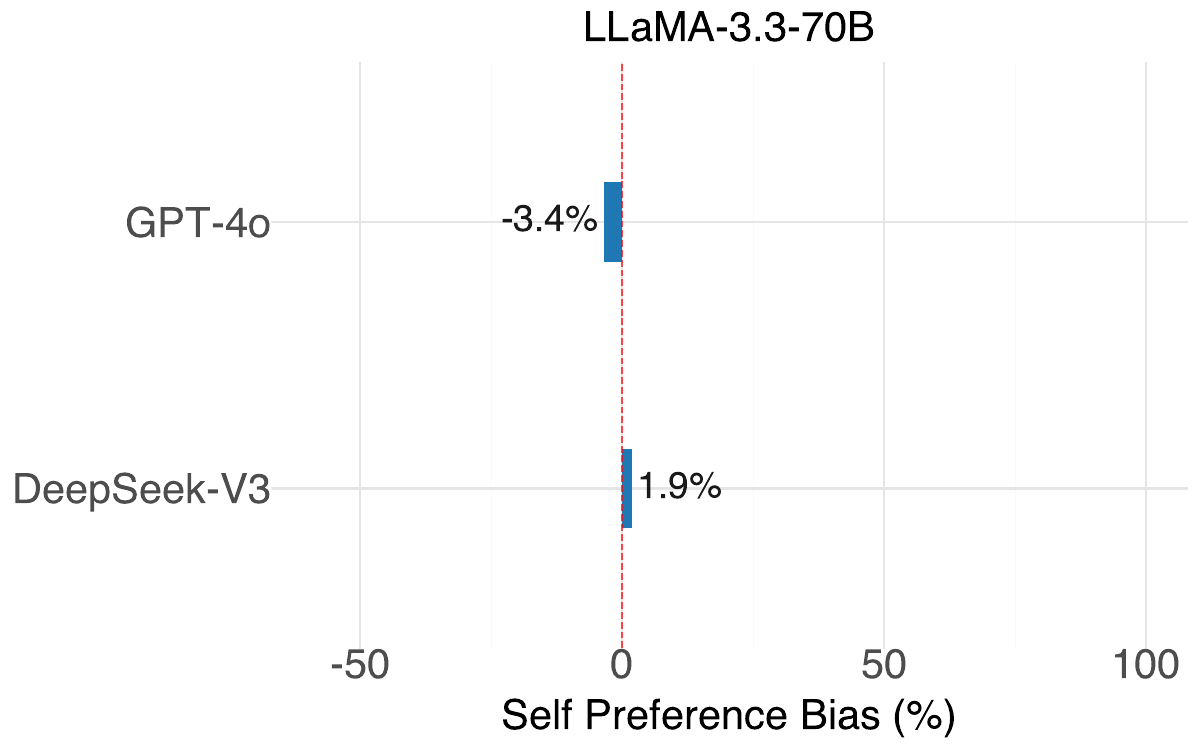}
  \caption{LLM-vs-LLM Equal Opportunity Self-preference Bias}
  \label{fig:eo-row}
\end{subfigure}
\caption{LLM-vs-LLM Self-Preference Bias}
\label{fig:llm-self-bias-combined}
\end{figure}

Analogous to the LLM-vs-Human setting, we assess statistical parity by comparing the probability that an evaluator LLM selects its own generated resume with the probability that it selects a resume produced by an alternative LLM, as defined in Eq.~(\ref{eq:AI_self_bias}). As shown in \autoref{fig:sp-row}, self-preferencing across LLMs is heterogeneous. DeepSeek-V3 is $69\%$ more likely to select its own resumes than those generated by LLaMA-3.3-70B, and $28\%$ more likely than those generated by GPT-4o. GPT-4o is $45\%$ more likely to select its own resumes than those produced by LLaMA-3.3-70B, but this tendency reverses against DeepSeek-V3, where it is $39\%$ more likely to select DeepSeek-V3's resumes over its own. LLaMA-3.3-70B shows only weak asymmetries: it is $1.6\%$ more likely to select its own resumes over GPT-4o's, but $6.7\%$ more likely to select DeepSeek-V3's resumes over its own.

\subsubsection{Equal Opportunity Self-Preference Bias} ~\

Similar to the LLM-vs-Human setting, we apply two complementary approaches to evaluate equal opportunity. First, we estimate conditional logistic regressions that control for content quality features. As shown in \autoref{fig:eo-row}, DeepSeek-V3 exhibits the strongest and most consistent self-preference bias among the three models, showing biases of $15.5\%$ against LLaMA 3.3-70B and $12\%$ against GPT-4o. By contrast, GPT-4o does not exhibit strong self-preference relative to LLaMA 3.3-70B and, in fact, shows a negative bias when compared with DeepSeek-V3, favoring DeepSeek-V3's outputs over its own. LLaMA 3.3-70B shows little self-preference in either comparison. Detailed conditional logistic regression results are provided in \autoref{tab:logistic_reg_2nd_bias}. 



Second, we benchmark model decisions against human judgments of resume quality.  These annotation-based results provide weaker evidence of equal opportunity self-preference. Although LLaMA 3.3-70B has a positive estimated bias of $12.94\%$ relative to GPT-4o, the estimate is not statistically significant ($95\%$ CI: [$-21.89\%$, $47.27\%$]). Similarly, neither DeepSeek-V3 nor GPT-4o exhibits statistically significant equal opportunity self-preference under the annotation-based benchmark, despite showing strong statistical parity self-preference bias in the previous analysis.

These findings highlight that LLM-vs-LLM self-preference is model-specific and generally weaker than LLM-vs-Human self-preference. The heterogeneity observed suggests that self-preference is not a uniform property of LLMs but may instead reflect model-specific factors, such as differences in stylistic alignment or the ability to recognize patterns in their own outputs.


\subsection{Robustness Checks}
In the following, we assess the robustness of our empirical results by considering an alternative approach to constructing counterfactual resumes in the correspondence experiment. In our baseline design, counterfactual summaries are generated from the remainder of the resume content. As a robustness check, we instead prompt each LLM to \emph{revise} the existing human-written executive summary by improving its clarity and wording, while explicitly preserving the candidate's experience and qualifications. This alternative construction mirrors a common and realistic use case of AI-assisted writing, in which job applicants employ LLMs to polish their resumes rather than generate summaries from scratch. This design allows us to examine whether AI self-preference persists under AI-assisted revision rather than reconstruction, thereby ruling out the possibility that our main results are driven by the specific method used to generate counterfactual text. Using this revised setup, we re-estimate both LLM-vs-Human and LLM-vs-LLM self-preference biases under the same evaluation framework, including conditional logistic regressions to assess equal opportunity bias. 

\begin{figure}[tb]
\centering
\begin{subfigure}[tb]{0.8\linewidth}
    \centering
    \includegraphics[width=\linewidth]{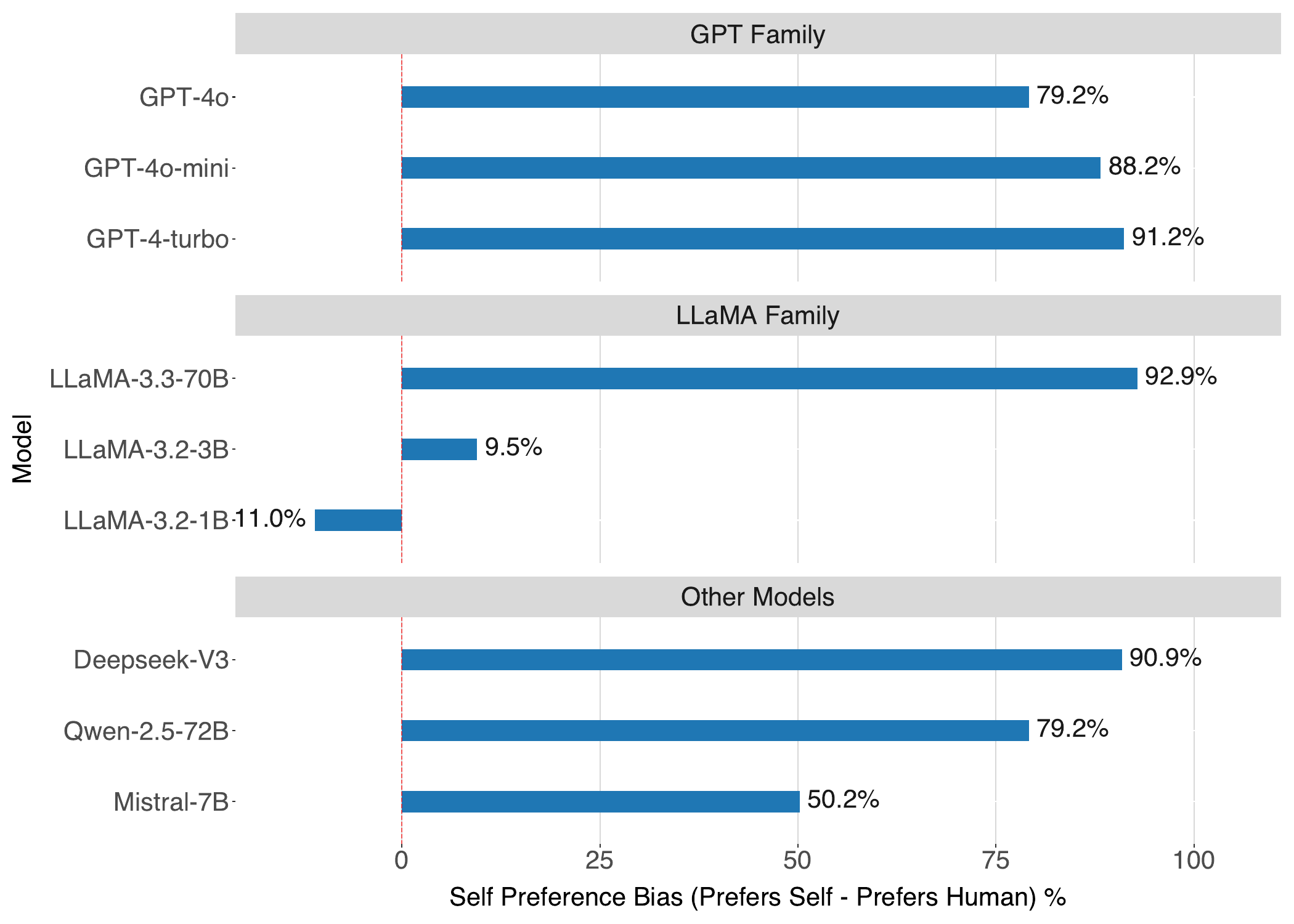}
    \caption{LLM-vs-Human Statistical Parity Self-preference Bias}
    \label{fig:ai-self-bias_revised}
\end{subfigure}
\vfill
\begin{subfigure}[tb]{0.8\linewidth}
    \centering
    \includegraphics[width=\linewidth]{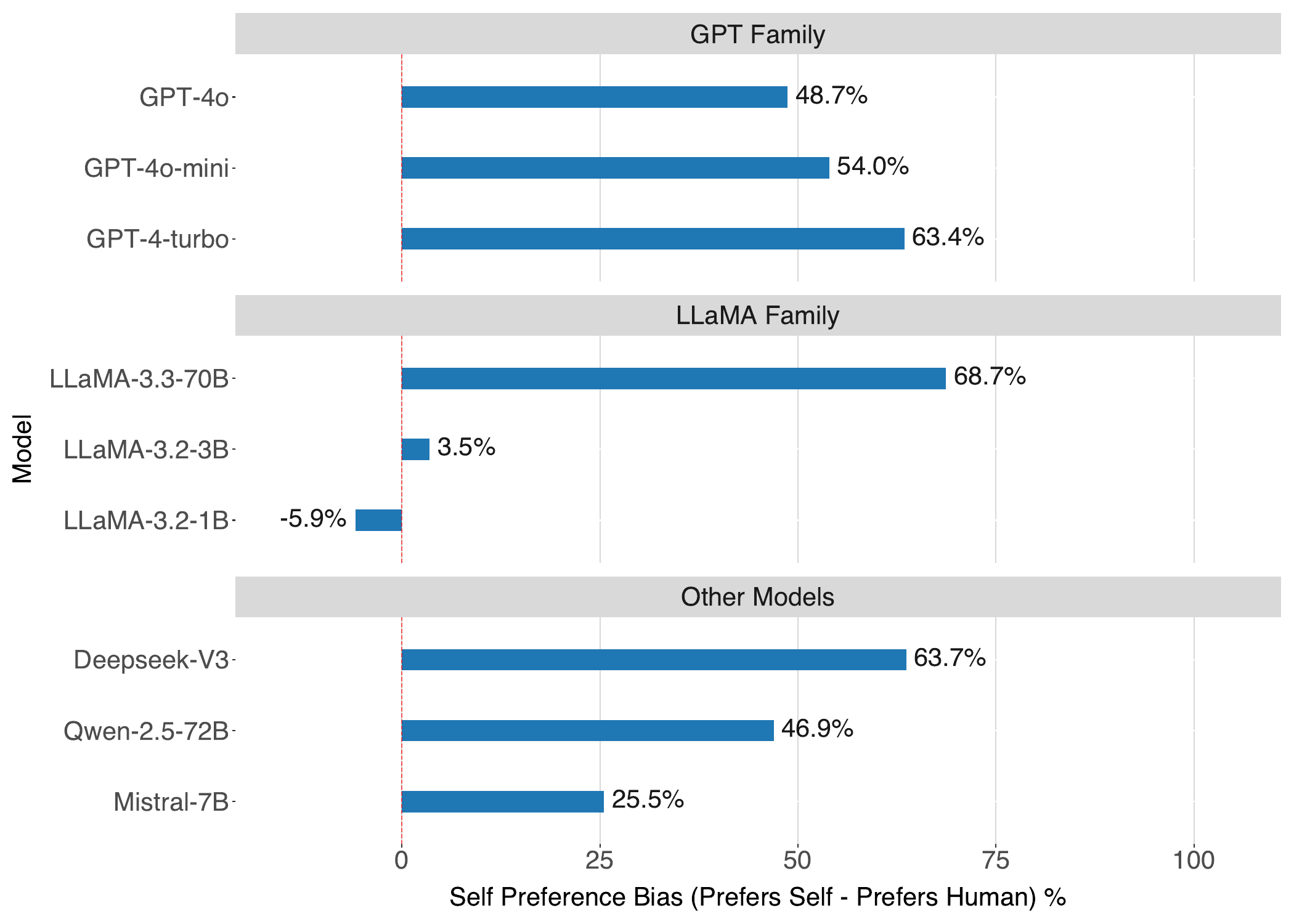}
    \caption{LLM-vs-Human Equal Opportunity Self-preference Bias}
    \label{fig:ai-self-EO-bias_revised}
\end{subfigure}
\caption{LLM-vs-Human Self-Preference Bias under Resume Revision Condition}
\label{fig:self-bias-combined_revised}
\end{figure}

As shown in \autoref{fig:ai-self-bias_revised} and \autoref{fig:ai-self-EO-bias_revised}, across both statistical parity and equal opportunity metrics, the qualitative patterns of LLM-vs-Human self-preference remain largely unchanged under the resume revision design. In both specifications, most evaluator LLMs continue to favor summaries revised by themselves over the original human-written summaries. Although the magnitude of self-preference is attenuated, for some models the magnitude still exceed $90\%$.
Such results suggest that LLM-vs-Human self-preference is a robust phenomenon that arises even when LLM-generated text is obtained through revision rather than generation from scratch. The corresponding conditional logistic regression estimates are reported in \autoref{tab:logistic_reg_revised}.

\begin{figure}[tb]
\centering

\begin{subfigure}[tb]{\textwidth}
  \centering
 \includegraphics[width=0.32\linewidth]{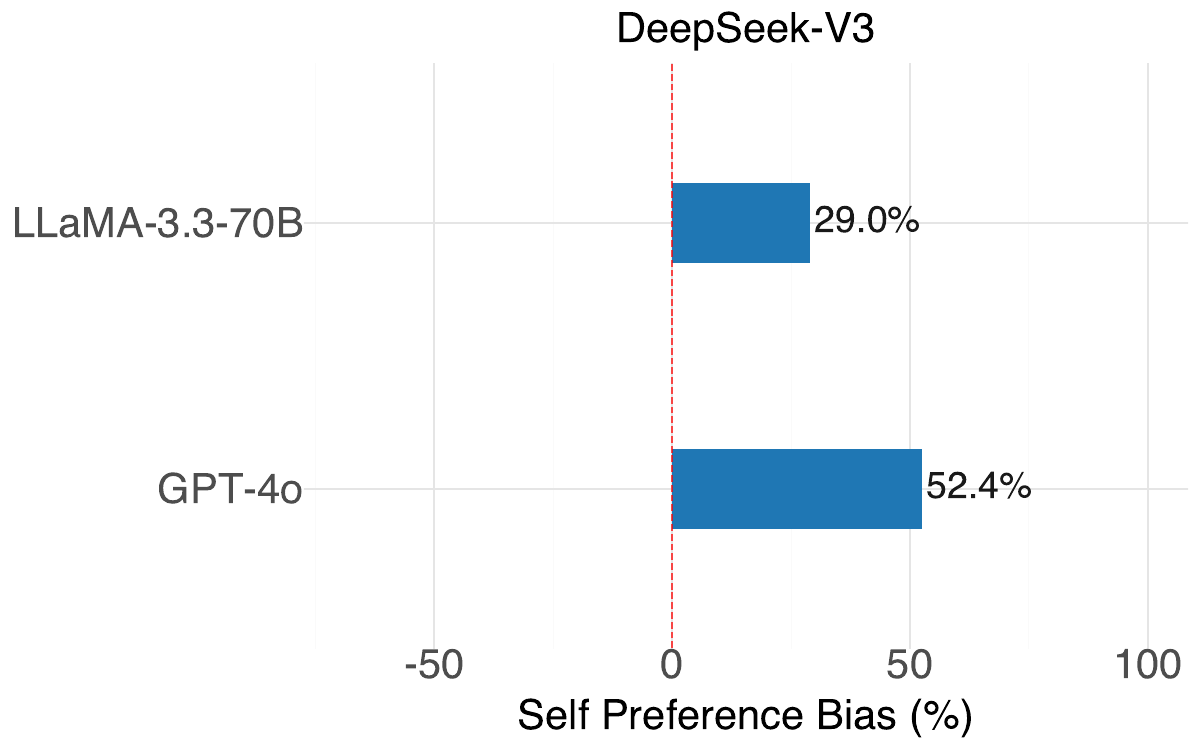}\hfill
  \includegraphics[width=0.32\linewidth]{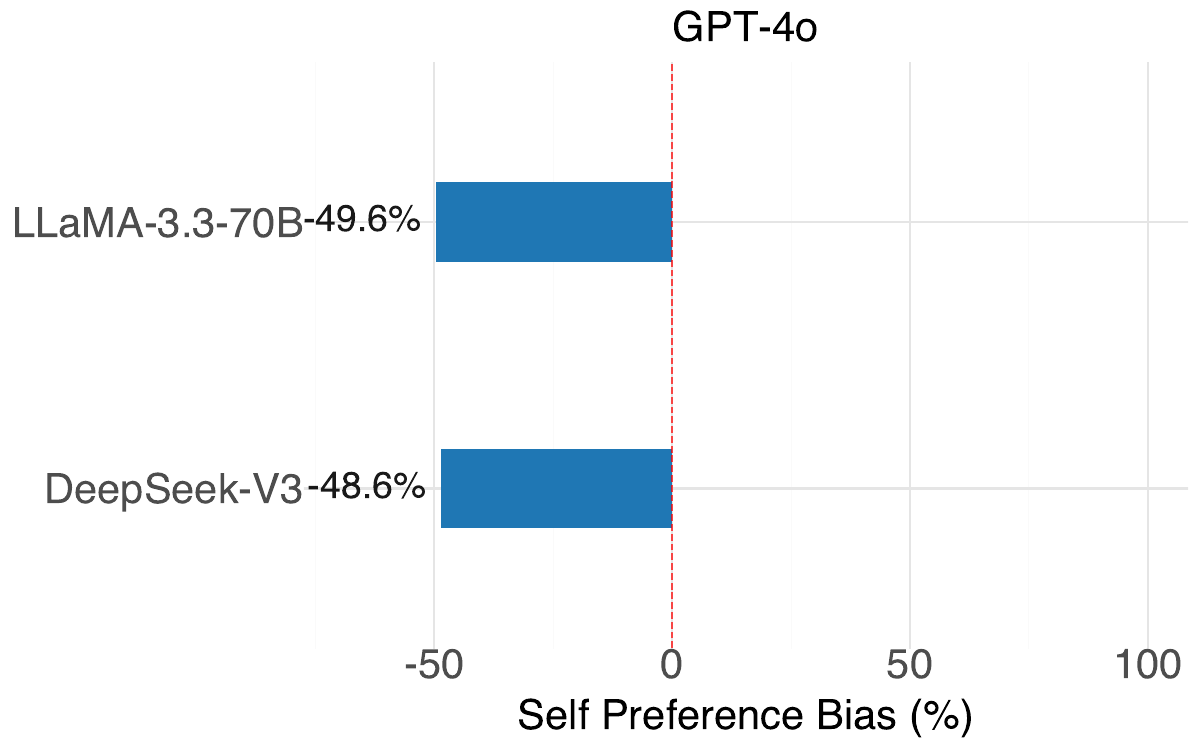}\hfill
  \includegraphics[width=0.32\linewidth]{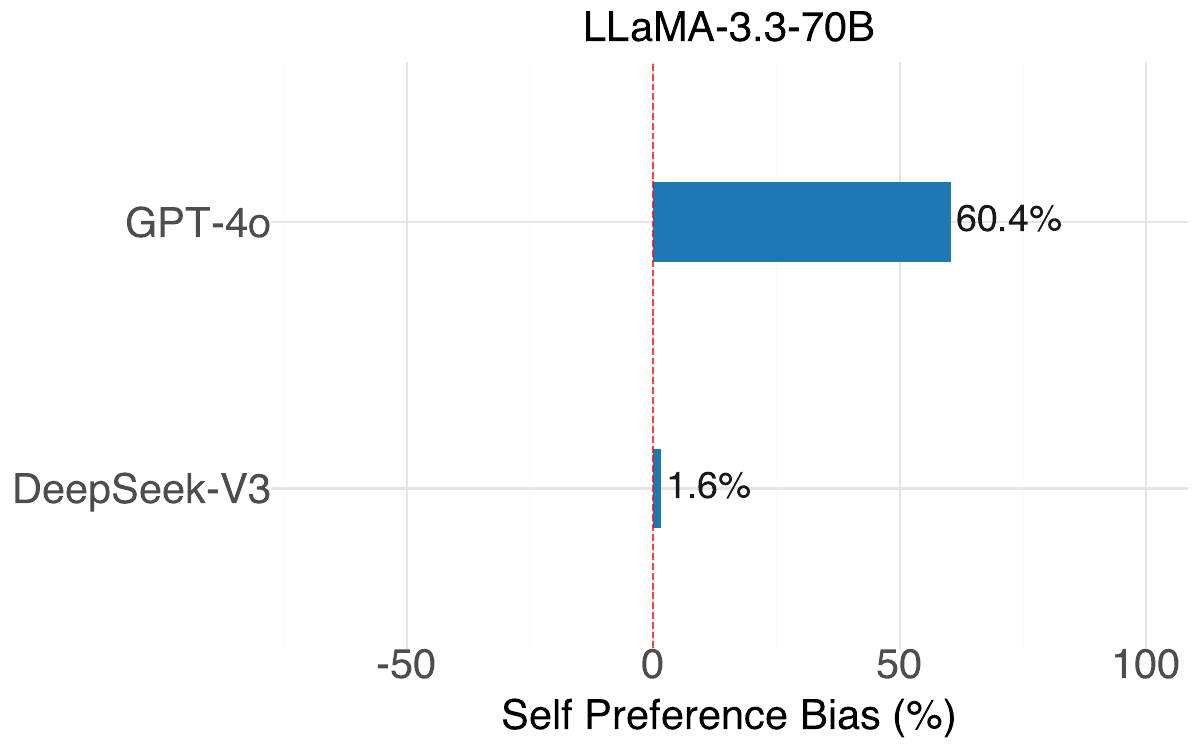}
  \caption{LLM-vs-LLM Statistical Parity Self-preference Bias}
  \label{fig:sp-row_revised}
\end{subfigure}

\vspace{2ex}

\begin{subfigure}[tb]{\textwidth}
  \centering
  \includegraphics[width=0.32\linewidth]{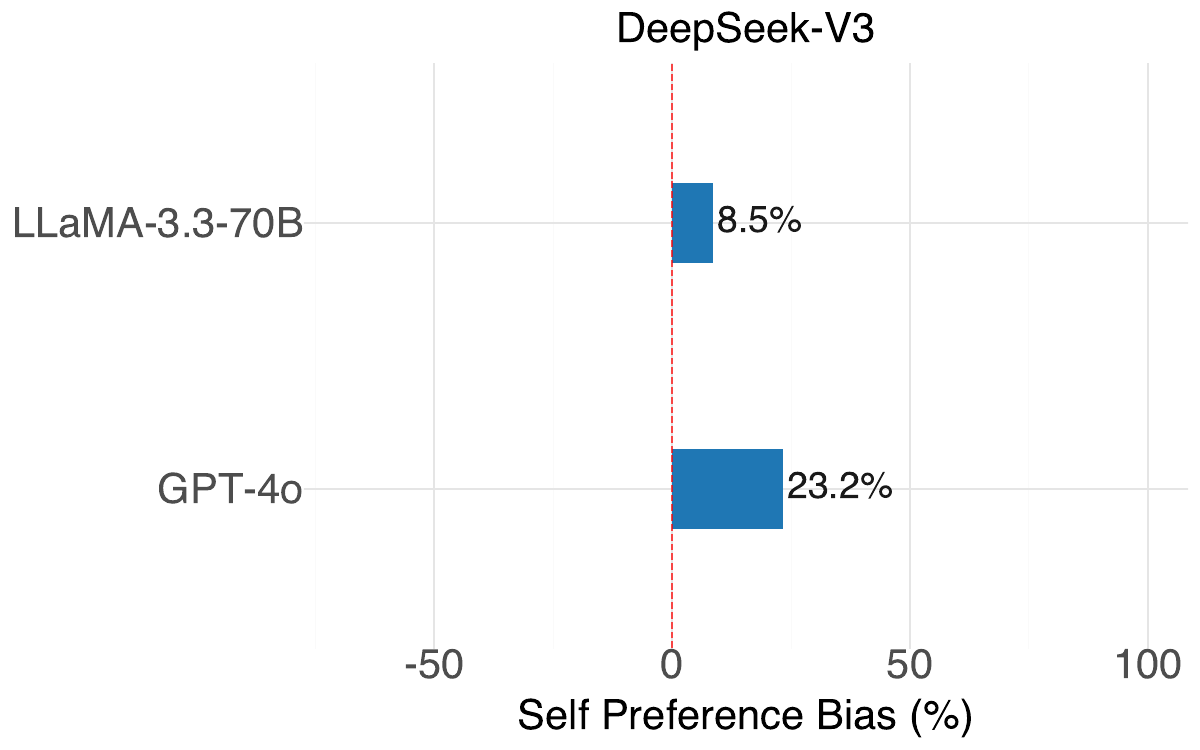}\hfill
  \includegraphics[width=0.32\linewidth]{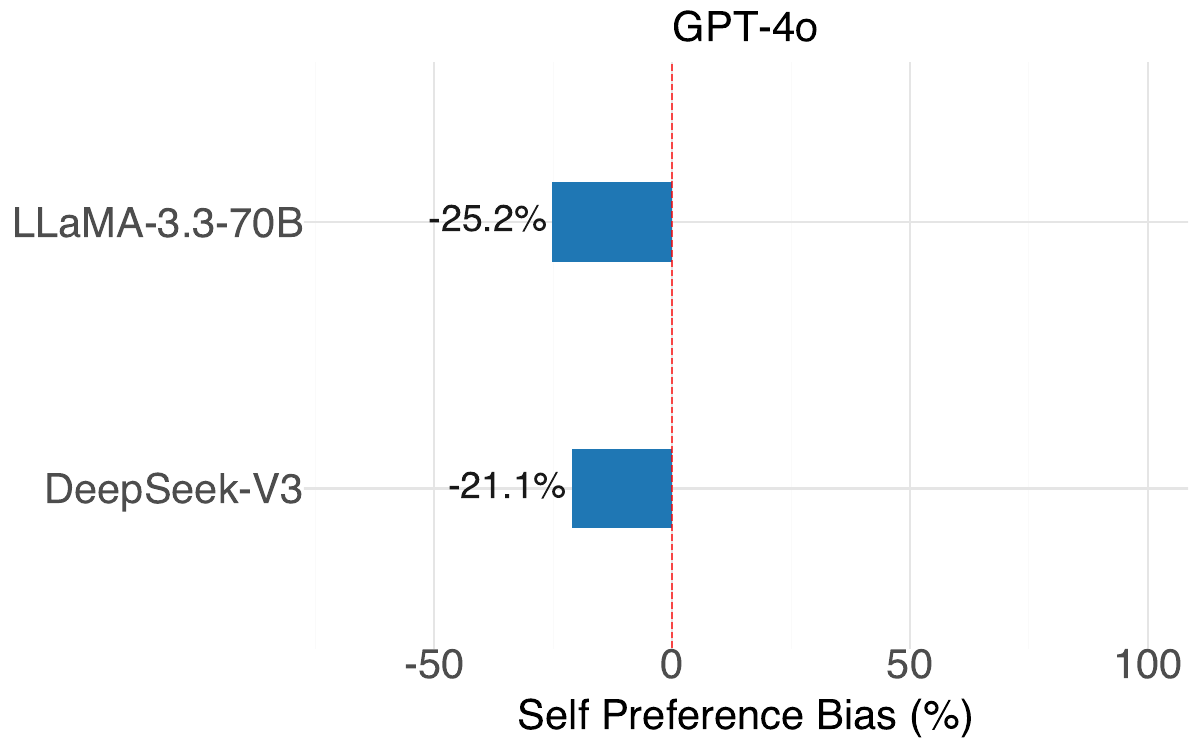}\hfill
  \includegraphics[width=0.32\linewidth]{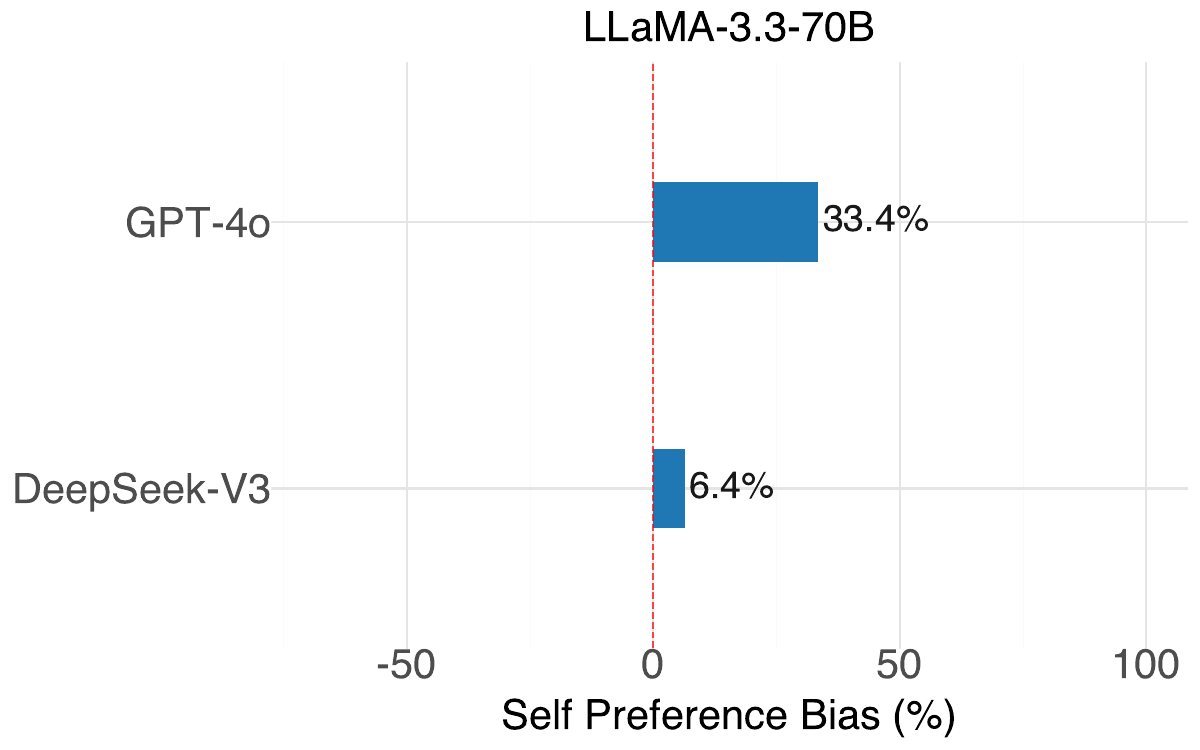}
  \caption{LLM-vs-LLM Equal Opportunity Self-preference Bias}
  \label{fig:eo-row_revised}
\end{subfigure}
\caption{LLM-vs-LLM Self-Preference Bias under Resume Revision Condition}
\label{fig:llm-self-bias-combined_revised}
\end{figure}

Turning to LLM-vs-LLM self-preference, the revision-based design yields more heterogeneous estimates than the baseline reconstruction specification (see \autoref{fig:sp-row_revised} and \autoref{fig:eo-row_revised}), but several qualitative patterns persist. Consistent with the conditional logistic regression results reported in \autoref{tab:logistic_reg_evaluator_revised}, DeepSeek-V3 continues to exhibit the strongest and most consistent equal-opportunity self-preference. Under the revision condition, DeepSeek-V3 displays equal-opportunity biases of approximately $23.2\%$ relative to GPT-4o and $8.5\%$ relative to LLaMA-3.3-70B---magnitudes that are comparable to, and in some cases larger than those observed under resume summary reconstruction. In addition, LLaMA-3.3-70B also exhibits substantial self-preference when evaluated against GPT-4o, indicating that asymmetric self-preference across model pairings persists even under the revision-based counterfactual.

\subsection{Impact of Self-Preference Bias in Algorithmic Hiring}
\begin{figure}[tb]
    \centering
        \centering
        \includegraphics[width=\linewidth]{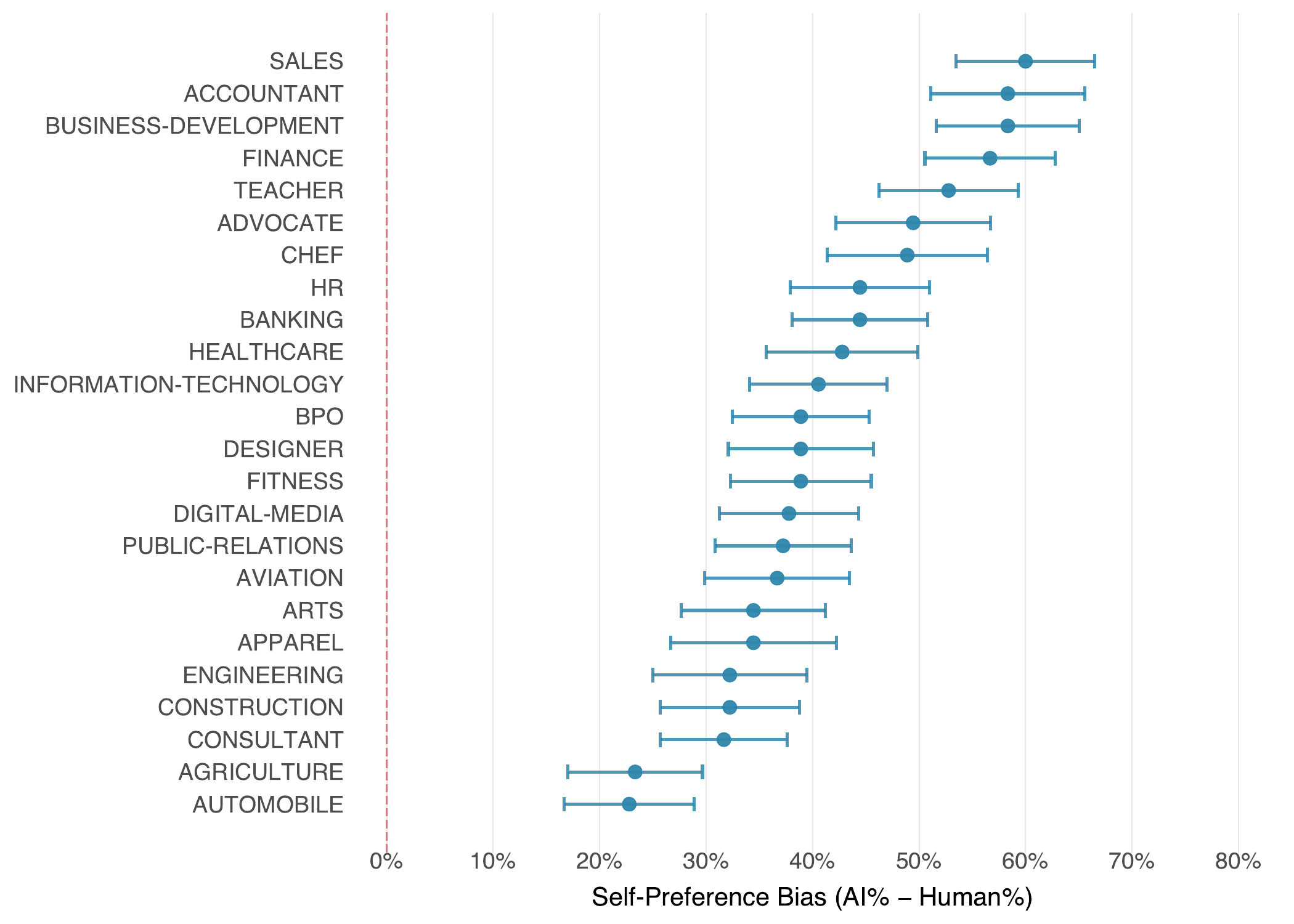}
   \caption{Self-Preference by Job Category} 
    \label{fig:bias_category}
    \begin{minipage}{\linewidth}
\vspace{1ex}
\footnotesize \textit{Notes:} Each bar shows the average self-preference bias across three evaluator LLMs (GPT-4o, DeepSeek-V3, and LLaMA-3.3-70B) in simulated hiring pipelines. Results are averaged across models; outcomes for individual models are reported in the \aref{appendix:bias_category}. Positive values indicate that candidates using the evaluator LLM are more likely to be shortlisted than those submitting human-written resumes. Across occupations, evaluator-generated resumes are consistently overrepresented among those selected.
\end{minipage}
\end{figure}
To quantify the practical implications of self-preference bias, we simulate a resume-screening pipeline modeled on competitive job markets. The simulation covers $24$ occupational categories, each with $30$ runs. In each run, we sample five candidate profiles and construct paired resumes: one human-written summary and one counterfactual summary generated by the evaluator LLM itself. These ten resumes (five human, five AI) form a candidate pool competing for four interview slots. The pool is randomly shuffled and presented to an evaluator model (GPT-4o, DeepSeek-V3, or LLaMA-3.3-70B), which is required to return exactly four finalists in ranked order. Since the AI-generated summaries are counterfactuals of the human-written ones, substantive content is held constant and we effectively control for resume quality. Thus, in the absence of bias, human and AI resumes should be equally likely to be selected---on average, two of each from every pool of ten candidates. For each run, we record the number of human versus AI resumes shortlisted, and then aggregate results across runs within each job category. This allows us to compute the magnitude of self-preference bias and construct $95\%$ confidence intervals at the category level.

The results shown in \autoref{fig:bias_category} reveal a systematic bias in favor of candidates who use the evaluator LLM to craft their resumes. In the absence of bias, we would expect resumes to be selected evenly, two human-written and two evaluator LLM-generated out of every four interview slots, so the probabilities in the figure would cluster around zero. Positive values indicate that evaluator LLM-generated resumes are more likely to be shortlisted, while values below zero would indicate the opposite. According to our results, however, all values lie well above zero, showing that evaluator-generated resumes are consistently overrepresented among those selected. On average across the three evaluator models, candidates using the evaluator LLM are $23\%$ to $60\%$ more likely 
to be shortlisted than those submitting human-written resumes. The disadvantage for human-written resumes is most pronounced in business-related occupations such as sales and accountant, and least evident in fields such as automobile and agriculture. Repeating the pipeline $30$ times per job category yields consistent results, with confidence intervals that exclude zero in every case. 

The consequences extend beyond individual selection outcomes. If access to certain LLMs is uneven across socioeconomic or linguistic groups, this bias risks amplifying existing inequities in job access. Over repeated hiring cycles, such dynamics can create a ``lock-in'' effect, where the stylistic patterns of the dominant LLM become entrenched in applicant pools and further reinforce its advantage. For employers, this presents a double-edged challenge: while LLM-based screening promises efficiency and a more comprehensive assessment than traditional keyword matching, it simultaneously increases the risk of overlooking highly qualified candidates who do not use the ``right'' AI tool, while advancing less-qualified candidates whose resumes happen to align stylistically with the evaluator.


\section{Bias Mitigation}
\label{sec:mitigation}


Having established both the prevalence and labor market impact of self-preference bias in LLM-based resume evaluations, we next turn to potential remedies. Left unaddressed, this bias can distort hiring outcomes by systematically advantaging candidates who use the same LLM as employers, while disadvantaging equally qualified applicants who do not. Such dynamics raise fairness concerns for job seekers and pose risks for employers, who may inadvertently overlook strong candidates. In practice, employers are likely to seek mitigation tools that are simple, cost-effective, and compatible with existing screening workflows. Motivated by evidence that self-preference bias is linked to LLM models' ability to recognize their own outputs, we evaluate two intervention strategies that directly target this self-recognition mechanism.

\subsection{Mechanism: Self-Recognition}
Recent work suggests that LLMs may possess an implicit ability to recognize text they have generated, and that this capacity is linked to self-preference. Benchmark datasets designed to probe situational awareness show that models can identify aspects of their own outputs and contexts \citep{laine2024me}. LLMs can also reliably distinguish their own generations from those of alternative models, with higher self-recognition capability often correlated with stronger self-preferencing biases \citep{Panickssery2024}. In addition, larger models appear to exhibit greater self-recognition, which may help explain why they show stronger self-preference in our empirical analysis \citep{laine2024me,Panickssery2024}. Taken together, these findings suggest that self-recognition is a plausible mechanism contributing to the bias we document. In the following, we evaluate mitigation strategies designed to directly target this mechanism.

\begin{table}[tbhp]
\centering
\caption{LLM-vs-Human Self-Preference Bias Before and After Mitigation}
\label{tab:self_preference_mitigation}
\begin{threeparttable}
\footnotesize
\begin{tabular}{lcccccc}
\toprule
\textbf{Bias Measure} & \textbf{GPT-4o} & \textbf{LLaMA-3.3-70B} & \textbf{DeepSeek-V3} \\
\midrule
\multicolumn{4}{l}{\textit{Before Mitigation}} \\
Self-Preference Bias (\%) & 82 & 79 & 72 \\
\midrule
\multicolumn{4}{l}{\textit{(1) After Mitigation via System Prompting}} \\
Self-Preference Bias (\%) & 61 & 30 & 60 \\
Absolute Decrease (pp) $\downarrow$ & 21 & 49 & 12 \\
Relative Decrease (\%) $\downarrow$ & 26 & 62 & 17 \\
\midrule
\multicolumn{4}{l}{\textit{(2) After Mitigation via Majority Voting}} \\
Self-Preference Bias (\%) & 30 & 23 & 29 \\
Absolute Decrease (pp) $\downarrow$ & 52 & 56 & 43 \\
Relative Decrease (\%) $\downarrow$ & 63 & 71 & 60 \\
\bottomrule

\end{tabular}
\vspace{0.5em}
\begin{minipage}{\linewidth}
\footnotesize \textit{Notes:} This table reports the LLM-vs-Human self-preference bias for three models before and after applying two mitigation strategies: (1) system prompting and (2) majority voting. Absolute decreases are reported in percentage points. Relative decreases are calculated as the percent reduction from the pre-mitigation bias. 
\end{minipage}
\end{threeparttable}
\end{table}


\subsection{System Prompting}
Our first mitigation strategy addresses self-preference by disrupting the evaluator's tendency to rely on stylistic or linguistic cues that signal its own outputs. Specifically, we modify the evaluator's system prompt to explicitly discourage source-based judgments and instead focus attention on substantive content quality. For example, the revised prompt instructs: ``You should not consider or infer whether the resumes were written by a human or by AI. Focus only on the quality of the content.'' This intervention aims to weaken the influence of self-recognition cues that drive models to favor their own generative style. 

Applying this prompting strategy across evaluator models leads to consistent reductions in self-preference bias. For example, GPT-4o's LLM-vs-Human bias decreases from $82\%$ to $61\%$, and LLaMA-3.3-70B's from $79\%$ to $30\%$, after controlling for resume quality (\autoref{tab:self_preference_mitigation}). These results indicate that the bias is not hardwired into model architecture but can be shaped by context and instruction, providing evidence that explicit prompts can partially disrupt self-recognition.

\subsection{Majority Voting Ensemble}
Our second strategy mitigates self-preference bias through ensemble evaluation. Instead of relying on a single LLM to judge a resume pair, we construct a panel of three models: the target evaluator and two smaller models (LLaMA-3.2-1B and LLaMA-3.2-3B) that exhibit minimal self-preference. The final decision is determined by majority vote. This approach is motivated by recent evidence that stronger self-recognition ability is associated with greater self-preference bias \citep{Panickssery2024}. By combining models with weaker self-recognition tendencies, the ensemble leverages model diversity to dilute the bias of larger evaluators.

This mitigation strategy proves highly effective. Across all three LLM models tested, the average LLM-vs-Human comparisons self-preference bias can drop by over $50\%$. For example, as it is shown in \autoref{tab:self_preference_mitigation}, GPT-4o's bias is reduced from $82\%$ to $30\%$, and LLaMA-3.3-70B's from $79\%$ to $23\%$. These results demonstrate that both system prompting and majority voting offer robust and scalable approaches to mitigation, particularly in high-stakes applications where fairness is critical.

In summary, the two mitigation strategies demonstrate that self-preference bias, while widespread, is not immutable. With relatively simple design interventions, we can substantially reduce bias in LLM-based evaluations without modifying the underlying model weights or retraining. These findings offer a practical path forward for deploying LLMs in evaluative roles while minimizing unintended algorithmic unfairness. 

\section{Concluding Remarks}
\label{section: concluding_remarks}
Our study documents a systematic form of algorithmic bias, AI self-preference, in the context of algorithmic hiring. Across two fairness metrics, we observe strong and consistent evidence of LLM-vs-Human self-preference in nearly all models tested. Simulation experiments show that in realistic hiring pipelines, candidates using the same LLM as the evaluator LLM are $23\%$ to $60\%$ more likely to be shortlisted than if they submit human-written resumes, with the largest disadvantage observed in business-related occupations such as sales and accounting. To address this issue, we propose two mitigation strategies (system prompting and majority voting) that firms can adopt with minimal implementation costs. Both approaches substantially reduce bias, cutting self-preference by more than half.

Beyond technical remedies, our findings have important policy implications. Current discussions of AI fairness largely focus on demographic disparities, but our results highlight the need to address biases arising from AI–AI interactions. Regulators and hiring platforms should recognize AI self-preference as a distinct and emerging form of algorithmic bias. Transparency requirements could mandate that organizations disclose whether AI is used in resume screening and what safeguards are in place to ensure fairness. In addition, third-party audits could incorporate self-preference metrics into fairness evaluations of AI-assisted hiring systems. Such measures would promote more accountable and equitable deployment of AI in employment contexts.

Finally, our study opens several directions for future research. As AI tools become more widely adopted, interactions between AIs will become increasingly common. Extra caution is warranted in contexts where AI systems are placed in evaluative or adjudicative roles, such as content moderation, grading in higher education, or other settings where they act as evaluator, judges or mediators. Another important direction is to examine how self-preference bias manifests in multilingual or cross-cultural environments, where non-English content may be especially vulnerable due to tokenization artifacts or limited representation in training data. Last but not the least, further investigation into the mechanisms underlying self-preference is needed. Rigorous study of self-recognition and other contributing processes will be critical to addressing the root causes of this bias and ensuring the fair integration of AI into hiring and other high-stakes decision domains.

\bibliographystyle{bib/informs2014.bst}
\bibliography{bib/reference.bib}

\ECSwitch

\renewcommand{\theHsection}{A\arabic{section}}
\counterwithin*{equation}{section}
\renewcommand\theequation{EC.\thesection.\arabic{equation}}

\begin{center}
    \large \textbf{Online Appendices}
\end{center}

\begin{APPENDICES}
\section{Sample Resume}
\label{appendix:resume_example}
\begin{figure}[tphb]
    \centering
    \label{fig:enter-label}
    \includegraphics[width=0.8\linewidth]{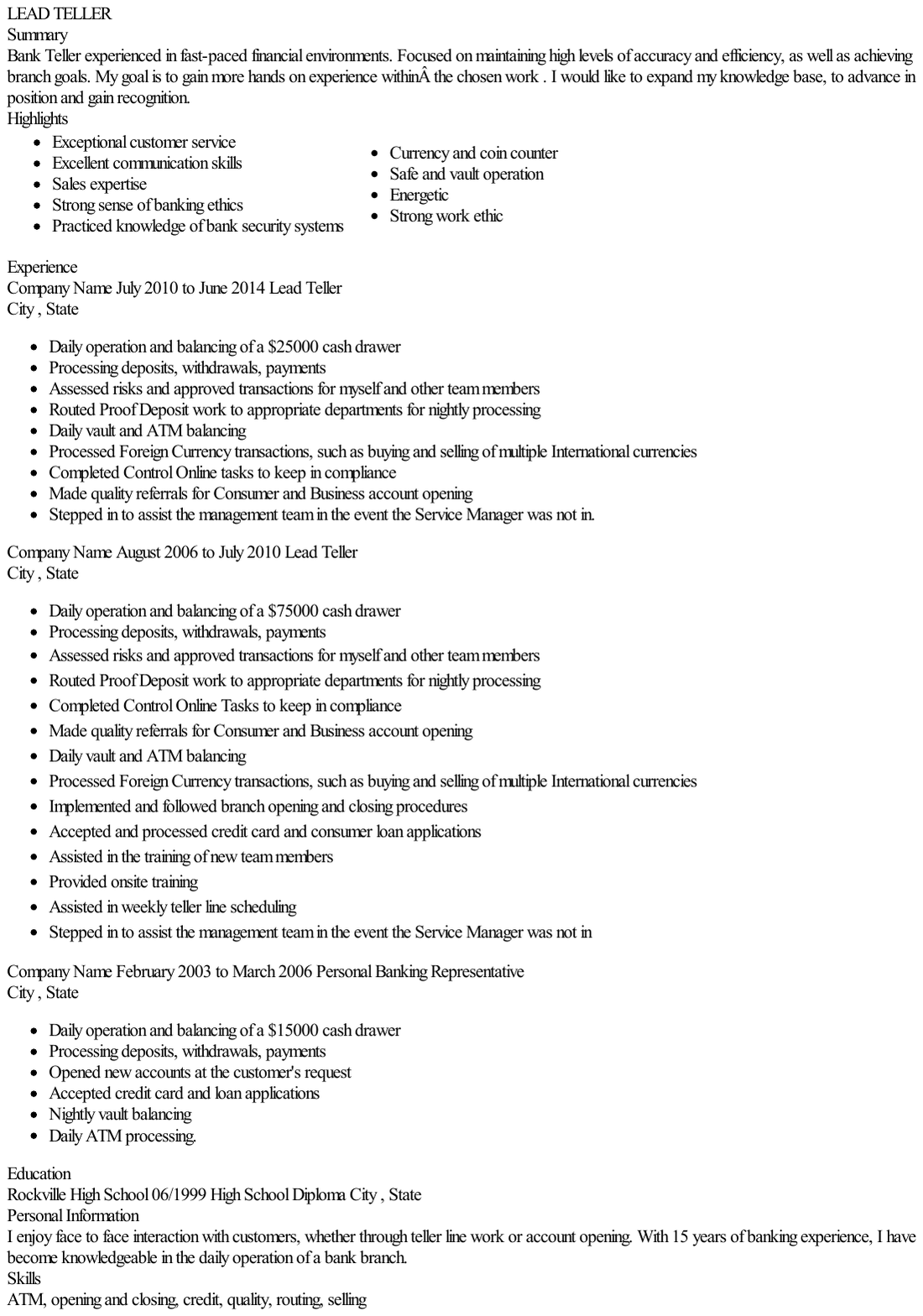}
        \caption{Sample Resume}
\end{figure}

\section{Prompts}
\label{appendix:prompts}
\subsection{Counterfactual Resume Generation}
\label{appendix:prompt_generation}
We use the following system prompt to instruct the LLMs for
the summary generation task:
\begin{quote}
\texttt{You are a highly skilled resume writer. Your task is to generate a concise, professional summary based on the provided resume.\\ 
The summary should be between 30 and 80 words, integrating seamlessly with the resume content.\\ 
Return only the summary as a single, well-structured sentence with no additional text.}
\end{quote}

The user prompt supplies the resume content in this format:
\begin{quote}
\texttt{Resume:\\
\{\textit{resume\_text}\}\\
Provide a resume summary (between 30 and 80 words) with no other text.}
\end{quote}
\subsection{Pairwise Resume Evaluation}
\label{appendix:prompt_evaluation}
\begin{quote}
The system prompt is
\texttt{You are an AI-powered resume screener designed to assist in candidate evaluation. Your role is to compare two candidate resumes and select the stronger one based on relevance, skills, and experience.
You respond with only "A" or "B" and no other text.}
\end{quote}
The user prompt for each comparison is structured as follows:
\begin{quote}
\texttt{Here are two candidate resumes for a job position. It is important for me to select the better resume for further consideration.\\
Resume A:\\
\{\textit{resume\_text\_A}\}\\
Resume B:\\
\{\textit{resume\_text\_B}\}\\
Can you tell me which candidate's resume is stronger? Please answer with only "A" or "B" and no other text.}
\end{quote}

\section{Sample Human-written and AI-generated Summaries}
\label{appendix:summary_example}

\begin{itemize}
    \item \textbf{Human}: Obtain a position working in customer service where I can utilize my friendly and outgoing attitude. I'm very a nalytical and adept at resolving complex network issues. Critical thinker who addresses customer support issues quickly and who consistently exceeds performance standards.
    \item \textbf{GPT-4-turbo}: A versatile professional with experience in education, healthcare, and customer service, proficient in Microsoft Office and fluent in English and Twi. Demonstrated excellence in high-stress environments with top customer satisfaction ratings and a strong ability to resolve issues effectively. Holds a Bachelor of Arts in African American Studies from The Ohio State University, with additional training in child development and business communications.
    \item \textbf{GPT-4o-mini}: Enthusiastic and organized educator with a Bachelor of Arts in African American Studies, recognized for exceptional customer service and teamwork, achieving over 80\% customer satisfaction. Proficient in Microsoft Office and experienced in high-pressure environments, demonstrating excellent verbal and written communication skills. Bilingual in English and Twi, with a strong commitment to fostering a healthy learning environment and active involvement in community service through church activities.
    \item \textbf{GPT-4o}: Energetic and organized professional with excellent analytical and communication skills, fluent in English and Twi, and proficient in Microsoft Office; recognized for exceptional customer satisfaction and teamwork, with diverse experience as a teacher, nursing aide, receptionist, and cook, holding a Bachelor of Arts in African American Studies from The Ohio State University, and actively engaged in church choir and event decoration activities.
    \item \textbf{LLaMA 3.2}: Dedicated and detail-oriented professional with excellent communication skills, offering experience in education, customer service, and administration, proficient in Microsoft Office suite and fluent in English and Twi, with a strong background in childcare, event planning, and team collaboration, driven to leverage skills for a dynamic role.
    \item \textbf{LLaMA 3.2-1B}: Proven teaching and customer service experience in high-volume, high-stress environments, driving exceptional results through strong analytical and verbal communication skills. Combining administrative expertise with passion for education and community engagement, I deliver results-driven solutions that exceed expectations.

\item \textbf{LLaMA 3.3-70B}: Results-driven and organized professional with teaching, customer service, and administrative experience, possessing excellent communication and analytical skills, with a strong track record of achieving high customer satisfaction and working effectively in high-volume environments.

\item \textbf{Mistral}:  An adaptable and customer-focused professional with a Bachelor's Degree in African American Studies, boasting experience in teaching, nursing, reception, and cooking roles. Strong verbal communication skills, proficient in Microsoft PowerPoint, Word, and Salesforce.com. Recipient of customer service recognition and adept at issue resolution. Fluent in English and Twi with a background in child development and cultural influences. Active in the community as a choir member, Sunday school teacher, children's choir director, and volunteer event decorator.

\item \textbf{Deepseek-V3}: Dedicated and versatile professional with a Bachelor of Arts in African American Studies, extensive experience in teaching, customer service, and healthcare support, and a proven ability to thrive in high-stress environments; skilled in verbal and written communication, Microsoft Office, and Salesforce.com, with a strong commitment to fostering positive relationships and maintaining organized, efficient operations.

\item \textbf{Qwen 2.5-72B}: Energetic and organized teacher with strong verbal and written communication skills, proficient in Microsoft Office, and fluent in English and Twi. Awarded for exceptional customer service and teamwork, with experience in education, healthcare, and front desk management, complemented by a Bachelor of Arts in African American Studies and extensive involvement in community and church activities.
\end{itemize}

\section{Human Annotation Interface}
\label{appendix:interface}
\begin{figure}[H]
    \centering
    \includegraphics[width=0.8\linewidth]{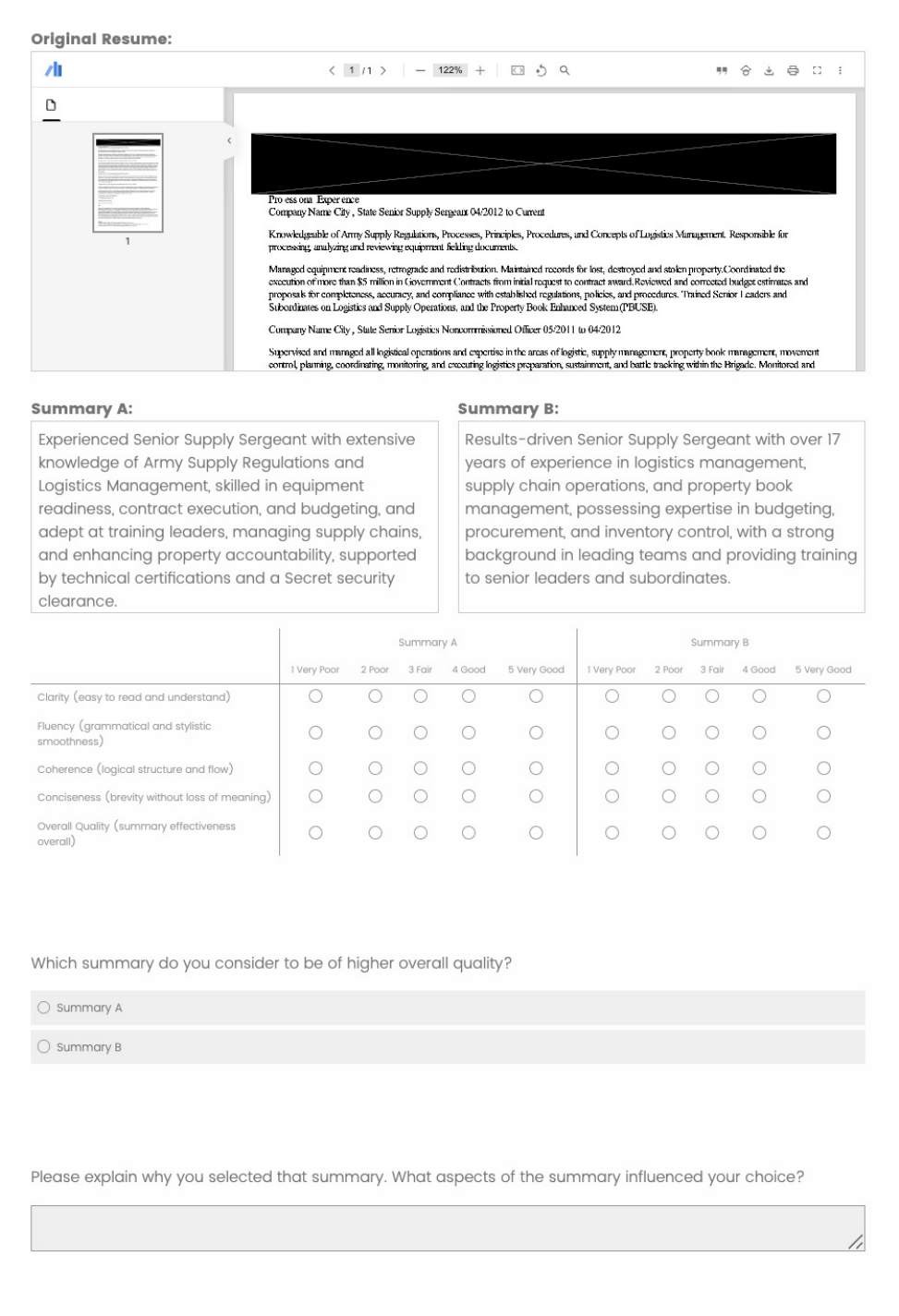}
    \label{fig:annotation-interface}
        \caption{Interface shown to human annotators for evaluating resume summaries. Annotators compare two summaries and indicate which one is of higher quality, without knowing the source (human or AI-generated).}
\end{figure}

\section{Annotation Instructions}
\label{appendix:instructions}

\subsection*{Objective}
The purpose of this study is to examine how people evaluate written resume summaries and to better understand the factors that influence perceptions of summary quality.

\subsection*{Task}
You will evaluate 32 pairs of resume summaries (2 of which contain attention checks). Each pair is based on the same original resume. Your task is to assess the quality of both summaries using the materials provided. Each question includes:
\begin{itemize}[noitemsep]
  \item A link to the original resume (PDF),
  \item Two corresponding summary versions.
\end{itemize}

\noindent For each pair, please complete the following:
\begin{enumerate}
  \item Rate each summary on five linguistic dimensions: \textit{clarity, fluency, coherence, conciseness}, and \textit{overall quality}.
  \item Select the better summary—the one you believe more effectively represents the original resume.
  \item \textit{(Optional but encouraged)}: Provide a brief rationale explaining your choice. Your feedback helps us understand how people assess resume quality.
\end{enumerate}

\subsection*{Incentive}
The study will take approximately \textbf{1 hour} to complete. You will receive \textbf{\$12} for your participation through Prolific. Additionally, the top 10\% of participants who provide the most persuasive and detailed free-text rationales will receive a \textbf{\$5 bonus}.

\subsection*{To Avoid Rejection}
\begin{itemize}
  \item Complete all 32 resume pairs fully.
  \item Provide thoughtful and consistent ratings—random or clearly careless responses will be excluded.
  \item Select a better summary for every pair (this is required).
  \item Engage with the content and do not leave required fields blank.
  \item Pass all attention checks.
\end{itemize}
We reserve the right to reject submissions that do not meet these quality standards.

\subsection*{Data Storage}
Your anonymized data will be stored securely for no more than two years. Personally identifiable information will not be shared outside the research team and will be destroyed after two years.



\subsection*{Risk and Benefits}
There are no known risks associated with participating in this research. While there are no direct benefits, participants will gain exposure to behavioral research methods and may benefit indirectly from the knowledge generated by this study.

\section{Conditional Logistic Regression Results}
\label{appendix:regression_results}
\begin{table}[H]
\centering
\footnotesize
\caption{Conditional Logistic Regression Results by Model Family}
\label{tab:logistic_reg_1st_bias}
\begin{threeparttable}
\centering
\vspace{0.5em}
\parbox{\linewidth}{
\centering
\begin{tabular}{l@{}c@{\hskip 6pt}c@{\hskip 6pt}c@{}}
\textbf{Panel A: GPT Models}\\
\toprule
& \multicolumn{3}{c}{Dependent Variable: $\textrm{Preferred}_{ij}$} \\
\cmidrule(lr){2-4}
Variables & GPT-4-turbo & GPT4o-mini & GPT4o \\
\midrule
$\textrm{evaluatorLLM}_{ij}$ & 1.616$^{***}$ & 1.653$^{***}$ & 2.305$^{***}$ \\
& (0.087) & (0.096) & (0.166) \\
\midrule
$\boldsymbol{\phi}_{ij}$: LIWC Features & Yes & Yes & Yes \\
$\boldsymbol{\psi}_{ij}$: Auto Scores & Yes & Yes & Yes \\
\midrule
Resume Pairs & 2245 & 2245 & 2245 \\
Observations & 4490 & 4490 & 4490 \\
Pseudo $R^2$ & 0.816 & 0.819 & 0.915 \\
Log Likelihood & -286.67 & -281.53 & -132.11 \\
\bottomrule\\
\textbf{Panel B: LLaMA Models} \\
\toprule
& \multicolumn{3}{c}{Dependent Variable: $\textrm{Preferred}_{ij}$} \\
\cmidrule(lr){2-4}
Variables & LLaMA-3.2-1B & LLaMA-3.2-3B & LLaMA-3.3-70B \\
\midrule
$\textrm{evaluatorLLM}_{ij}$ & -0.028 & 0.233$^{***}$ & 2.136$^{***}$ \\
& (0.030) & (0.035) & (0.152) \\
\midrule
$\boldsymbol{\phi}_{ij}$: LIWC Features & Yes & Yes & Yes \\
$\boldsymbol{\psi}_{ij}$: Auto Scores & Yes & Yes & Yes \\
\midrule
Resume Pairs & 2245 & 2245 & 2245 \\
Observations & 4490 & 4490 & 4490 \\
Pseudo $R^2$ & 0.005 & 0.059 & 0.892 \\
Log Likelihood & -1547.70 & -1464.50 & -168.41 \\
\bottomrule\\
\textbf{Panel C: Other Open-Source Models} \\
\toprule
& \multicolumn{3}{c}{Dependent Variable: $\textrm{Preferred}_{ij}$} \\
\cmidrule(lr){2-4}
Variables & Mistral-7B & Qwen-2.5-72B &  DeepSeek-V3\\
\midrule
$\textrm{evaluatorLLM}_{ij}$ & 0.575$^{***}$ & 2.092$^{***}$ & 1.797$^{***}$ \\
& (0.041) & (0.124) & (0.117) \\
\midrule
$\boldsymbol{\phi}_{ij}$: LIWC Features & Yes & Yes & Yes \\
$\boldsymbol{\psi}_{ij}$: Auto Scores & Yes & Yes & Yes \\
\midrule
Resume Pairs & 2245 & 2245 & 2245 \\
Observations & 4490 & 4490 & 4490 \\
Pseudo $R^2$ & 0.293 & 0.873 & 0.865 \\
Log Likelihood & -1100.90 & -198.05 & -209.52 \\
\bottomrule
\end{tabular}
}

\vspace{0.5em}
\begin{minipage}{0.9\linewidth}
\footnotesize \textit{Notes:} $^{***} p<0.01$, $^{**} p<0.05$, $^{*} p<0.1$. The $\textrm{evaluatorLLM}_{ij}$ coefficient represents each evaluator LLMs preference for its own outputs. 
Standard errors are reported in parentheses. Each model uses $2,245$ paired resume comparisons.
\end{minipage}
\end{threeparttable}
\end{table}

\begin{table}[H]
\centering
\footnotesize
\caption{Conditional Logistic Regression Results by Evaluators}
\label{tab:logistic_reg_2nd_bias}
\begin{threeparttable}
\vspace{0.5em}
\centering
\begin{tabular}{l@{}c@{\hskip 6pt}c@{}}
\textbf{Panel A: DeepSeek-V3 as Evaluator} \\
\toprule
& \multicolumn{2}{c}{Dependent Variable: $\textrm{Preferred}_{ij}$} \\
\cmidrule(lr){2-3}
Variables & GPT-4o & LLaMA-3.3-70B \\
\midrule
$\textrm{evaluatorLLM}_{ij}$  & 0.242$^{***}$   &  0.313$^{***}$ \\
& (0.028)  &  (0.105) \\
\midrule
$\boldsymbol{\phi}_{ij}$: LIWC Features & Yes & Yes  \\
$\boldsymbol{\psi}_{ij}$: Automatic Scores & Yes & Yes \\
\midrule
Resume Pairs & 2245 & 2245 \\
Observations & 4490 & 4490 \\
Pseudo $R^2$ & 0.064  &  0.397\\
Log Likelihood & -1455.90  &  -938.54   \\
\bottomrule \\
\textbf{Panel B: GPT-4o as Evaluator} \\
\toprule
& \multicolumn{2}{c}{Dependent Variable: $\textrm{Preferred}_{ij}$} \\
\cmidrule(lr){2-3}
Variables & LLaMA-3.3-70B & DeepSeek-V3 \\
\midrule
$\textrm{evaluatorLLM}_{ij}$ & 0.099  &   -0.361$^{***}$   \\
& (0.084) & (0.029)  \\
\midrule
$\boldsymbol{\phi}_{ij}$: LIWC Features & Yes & Yes  \\
$\boldsymbol{\psi}_{ij}$: Automatic Scores & Yes & Yes \\
\midrule
Resume Pairs & 2245 & 2245  \\
Observations & 4490 & 4490 \\
Pseudo $R^2$ & 0.178 & 0.134 \\
Log Likelihood & -1279.70 &  -1348.00\\
\bottomrule \\
\textbf{Panel C: LLaMA-3.3-70B as Evaluator} \\
\toprule
& \multicolumn{2}{c}{Dependent Variable: $\textrm{Preferred}_{ij}$} \\
\cmidrule(lr){2-3}
Variables & GPT-4o & DeepSeek-V3  \\
\midrule
$\textrm{evaluatorLLM}_{ij}$ & -0.068&  0.038    \\
& (0.075) &  (0.080)\\
\midrule
$\boldsymbol{\phi}_{ij}$: LIWC Features & Yes & Yes  \\
$\boldsymbol{\psi}_{ij}$: Automatic Scores & Yes & Yes  \\
\midrule
Resume Pairs & 2245 & 2245 \\
Observations & 4490 & 4490 \\
Pseudo $R^2$ & 0.004 &  0.010 \\
Log Likelihood &  -1550.00 &  -1540.80  \\
\bottomrule
\end{tabular}
\vspace{0.5em}
\begin{minipage}{\linewidth}
\vspace{0.5em}
\footnotesize \textit{Notes:} $^{***} p<0.01$, $^{**} p<0.05$, $^{*} p<0.1$. The $\textrm{evaluatorLLM}_{ij}$ coefficient represents each evaluator LLM's preference for its own outputs over other LLMs' outputs. Standard errors are reported in parentheses.
\end{minipage}
\end{threeparttable}
\end{table}



\newpage
\section{Robustness Checks}
\label{sec:robustness_check}
\subsection{Prompt}
\label{appendix:prompt_generation}
We use the following system prompt to instruct the LLMs to revise the resume. 
\begin{quote}
\texttt{You are helping a job candidate refine their resume summary. Your goal is to improve clarity and wording while keeping the candidate's experience and qualifications the same.}
\end{quote}
The user prompt is given below:
\begin{quote}
\texttt{Please revise the following resume summary to sound clearer and more polished, as if a job candidate used an AI tool to improve their resume. \\
Guidelines: \\
- Keep all experience, qualifications, and claims the same.\\
- Do NOT add new skills, roles, achievements, or responsibilities.\\
- Do NOT remove any existing information.\\
- Preserve all job titles, company names, dates, and numbers exactly.\\
- Do not exaggerate or downplay seniority, scope, or impact.\\
- Keep the overall length and structure similar to the original.\\
- Focus on improving flow, wording, and readability rather than content.\\
- Only output the revised summary, no other text.\\
Original resume summary:\\
\{\textit{HUMAN\_RESUME\_SUMMARY}\}\\
Rewritten resume summary:}
\end{quote}

\subsection{Sample Human-written and AI-revised Summaries}

\begin{itemize}
    \item \textbf{Human}: Dedicated, Driven, and Dynamic with over 20 years of customer service expertise. Motivated to maintain customer satisfaction and contribute to company success with an emphasis in personnel and process management. Solid team player with proven ability to establish rapport with clients.
    \item \textbf{GPT-4-turbo}: Committed and results-oriented professional with over 20 years of expertise in customer service. Highly motivated to enhance customer satisfaction and contribute to company success, specializing in personnel and process management. Proven team player with a strong ability to build relationships with clients.
    \item \textbf{GPT-4o-mini}: Dedicated, driven, and dynamic professional with over 20 years of expertise in customer service. Committed to enhancing customer satisfaction and supporting company success through a strong focus on personnel and process management. A solid team player with a proven ability to build rapport with clients.
    \item \textbf{GPT-4o}: A dedicated, driven, and dynamic professional with over 20 years of experience in customer service. Highly motivated to enhance customer satisfaction and contribute to company success, with a strong focus on personnel and process management. A solid team player, skilled in establishing rapport with clients.
    \item \textbf{LLaMA 3.2}: Results-driven customer service professional with over 20 years of experience driving customer satisfaction and company success through strategic personnel and process management. Leveraging a strong team-oriented approach, I foster meaningful relationships with clients by building rapport, ensuring seamless interactions, and cultivating long-term relationships that support business objectives.
    \item \textbf{LLaMA 3.2-1B}: Results-driven and customer-focused professional with 20+ years of experience in delivering exceptional customer service, driving business growth, and optimizing processes. Proven track record of building strong relationships with clients and colleagues, with a strong emphasis on delivering high-quality results and exceeding expectations.

\item \textbf{LLaMA 3.3-70B}: Results-driven professional with over 20 years of customer service expertise, dedicated to delivering exceptional customer satisfaction and driving company success through effective personnel and process management. As a collaborative team player, I possess a proven ability to build strong rapport with clients, fostering a positive and productive experience.

\item \textbf{Mistral}: A seasoned, driven, and dynamic professional with over 20 years of customer service experience. Committed to maintaining high levels of customer satisfaction and contributing to the overall success of organizations, with a focus on effective personnel and process management. Demonstrated ability to build strong relationships with clients, making me a valuable team player.

\item \textbf{Deepseek-V3}: Results-driven professional with over 20 years of customer service experience, specializing in personnel and process management to enhance customer satisfaction and drive organizational success. Skilled team collaborator with a proven track record of building strong client relationships.

\item \textbf{Qwen 2.5-72B}: Dedicated, driven, and dynamic with over 20 years of customer service expertise. Committed to maintaining customer satisfaction and contributing to company success through effective personnel and process management. Proven ability to build strong rapport with clients and excel as a solid team player.
\end{itemize}

\newpage
\subsection{Results}
\begin{table}[H]
\centering
\footnotesize
\caption{Conditional Logistic Regression Results by Model Family (Revised Summaries)}
\label{tab:logistic_reg_revised}
\begin{threeparttable}
\centering
\vspace{0.5em}
\parbox{\linewidth}{
\centering
\begin{tabular}{l@{}c@{\hskip 6pt}c@{\hskip 6pt}c@{}}
\textbf{Panel A: GPT Models}\\
\toprule
& \multicolumn{3}{c}{Dependent Variable: $\textrm{Preferred}_{ij}$} \\
\cmidrule(lr){2-4}
Variables & GPT-4-turbo & GPT4o-mini & GPT4o \\
\midrule
$\textrm{evaluatorLLM}_{ij}$ & 1.498$^{***}$ & 1.208$^{***}$ & 1.065$^{***}$ \\
& (0.076) & (0.060) & (0.047) \\
\midrule
$\boldsymbol{\phi}_{ij}$: LIWC Features & Yes & Yes & Yes \\
$\boldsymbol{\psi}_{ij}$: Auto Scores & Yes & Yes & Yes \\
\midrule
Resume Pairs & 2245 & 2245 & 2245 \\
Observations & 4490 & 4490 & 4490 \\
Pseudo $R^2$ & 0.753 & 0.700 & 0.535 \\
Log Likelihood & -384.82 & -467.17 & -724.22 \\
\bottomrule\\
\textbf{Panel B: LLaMA Models} \\
\toprule
& \multicolumn{3}{c}{Dependent Variable: $\textrm{Preferred}_{ij}$} \\
\cmidrule(lr){2-4}
Variables & LLaMA-3.2-1B & LLaMA-3.2-3B & LLaMA-3.3-70B \\
\midrule
$\textrm{evaluatorLLM}_{ij}$ & -0.118$^{***}$ & 0.070$^{**}$ & 1.685$^{***}$ \\
& (0.033) & (0.033) & (0.085) \\
\midrule
$\boldsymbol{\phi}_{ij}$: LIWC Features & Yes & Yes & Yes \\
$\boldsymbol{\psi}_{ij}$: Auto Scores & Yes & Yes & Yes \\
\midrule
Resume Pairs & 2245 & 2245 & 2245 \\
Observations & 4490 & 4490 & 4490 \\
Pseudo $R^2$ & 0.021 & 0.012 & 0.805 \\
Log Likelihood & -1523.53 & -1537.51 & -303.17 \\
\bottomrule\\
\textbf{Panel C: Other Open-Source Models} \\
\toprule
& \multicolumn{3}{c}{Dependent Variable: $\textrm{Preferred}_{ij}$} \\
\cmidrule(lr){2-4}
Variables & Mistral-7B & Qwen-2.5-72B &  DeepSeek-V3\\
\midrule
$\textrm{evaluatorLLM}_{ij}$ & 0.522$^{***}$ & 1.019$^{***}$ & 1.506$^{***}$ \\
& (0.035) & (0.046) & (0.068) \\
\midrule
$\boldsymbol{\phi}_{ij}$: LIWC Features & Yes & Yes & Yes \\
$\boldsymbol{\psi}_{ij}$: Auto Scores & Yes & Yes & Yes \\
\midrule
Resume Pairs & 2245 & 2245 & 2245 \\
Observations & 4490 & 4490 & 4490 \\
Pseudo $R^2$ & 0.197 & 0.527 & 0.743 \\
Log Likelihood & -1250.12 & -736.71 & -400.47 \\
\bottomrule
\end{tabular}
}

\vspace{0.5em}
\begin{minipage}{0.9\linewidth}
\footnotesize \textit{Notes:} $^{***} p<0.01$, $^{**} p<0.05$, $^{*} p<0.1$. The $\textrm{evaluatorLLM}_{ij}$ coefficient represents each evaluator LLMs preference for its own outputs (revised summaries vs. human original summaries). Standard errors are reported in parentheses.
\end{minipage}

\end{threeparttable}
\end{table}

\begin{table}[H]
\centering
\footnotesize
\caption{Conditional Logistic Regression Results by Evaluators (Revised Summaries)}
\label{tab:logistic_reg_evaluator_revised}
\begin{threeparttable}
\vspace{0.5em}
\centering
\begin{tabular}{l@{}c@{\hskip 6pt}c@{}}
\textbf{Panel A: DeepSeek-V3 as Evaluator} \\
\toprule
& \multicolumn{2}{c}{Dependent Variable: $\textrm{Preferred}_{ij}$} \\
\cmidrule(lr){2-3}
Variables & GPT-4o & LLaMA-3.3-70B \\
\midrule
$\textrm{evaluatorLLM}_{ij}$  & 0.472$^{***}$   &  0.170$^{***}$ \\
& (0.031)  &  (0.032) \\
\midrule
$\boldsymbol{\phi}_{ij}$: LIWC Features & Yes & Yes  \\
$\boldsymbol{\psi}_{ij}$: Automatic Scores & Yes & Yes \\
\midrule
Resume Pairs & 2245 & 2245 \\
Observations & 4490 & 4490 \\
Pseudo $R^2$ & 0.234  &  0.096\\
Log Likelihood & -1191.84  &  -1407.18   \\
\bottomrule \\
\textbf{Panel B: GPT-4o as Evaluator} \\
\toprule
& \multicolumn{2}{c}{Dependent Variable: $\textrm{Preferred}_{ij}$} \\
\cmidrule(lr){2-3}
Variables & LLaMA-3.3-70B & DeepSeek-V3 \\
\midrule
$\textrm{evaluatorLLM}_{ij}$ & -0.515$^{***}$  &   -0.428$^{***}$   \\
& (0.029) & (0.031)  \\
\midrule
$\boldsymbol{\phi}_{ij}$: LIWC Features & Yes & Yes  \\
$\boldsymbol{\psi}_{ij}$: Automatic Scores & Yes & Yes \\
\midrule
Resume Pairs & 2245 & 2245  \\
Observations & 4490 & 4490 \\
Pseudo $R^2$ & 0.208 & 0.237 \\
Log Likelihood & -1231.78 &  -1187.07\\
\bottomrule \\
\textbf{Panel C: LLaMA-3.3-70B as Evaluator} \\
\toprule
& \multicolumn{2}{c}{Dependent Variable: $\textrm{Preferred}_{ij}$} \\
\cmidrule(lr){2-3}
Variables & GPT-4o & DeepSeek-V3  \\
\midrule
$\textrm{evaluatorLLM}_{ij}$ & 0.695$^{***}$&  0.127$^{***}$    \\
& (0.031) &  (0.032)\\
\midrule
$\boldsymbol{\phi}_{ij}$: LIWC Features & Yes & Yes  \\
$\boldsymbol{\psi}_{ij}$: Automatic Scores & Yes & Yes  \\
\midrule
Resume Pairs & 2245 & 2245 \\
Observations & 4490 & 4490 \\
Pseudo $R^2$ & 0.305 &  0.060 \\
Log Likelihood &  -1081.33 &  -1463.02  \\
\bottomrule
\end{tabular}
\vspace{0.5em}
\begin{minipage}{\linewidth}
\vspace{0.5em}
\footnotesize \textit{Notes:} $^{***} p<0.01$, $^{**} p<0.05$, $^{*} p<0.1$. The $\textrm{evaluatorLLM}_{ij}$ coefficient represents each evaluator LLM's preference for its own revised outputs over other LLMs' revised outputs. Standard errors are reported in parentheses.
\end{minipage}
\end{threeparttable}
\end{table}

\section{Simulation Details and Results}
\label{appendix:bias_category}
\subsection*{Simulation Prompt}
We use the following prompt to simulate job-market screening:
\begin{quote}
\texttt{You are an AI-powered resume screener tasked with assisting in candidate evaluation. 
You will be given ten candidate resume summaries. 
Your job is to review their skills and experience, then select exactly four candidates who are the best fit for the role. 
Respond only with the candidate IDs of the four selected candidates, listed in order of preference, separated by commas and no other text.}
\end{quote}
\subsection*{Simulation Results}
\begin{figure}[tb]
    \centering
        \centering
        \includegraphics[width=\linewidth]{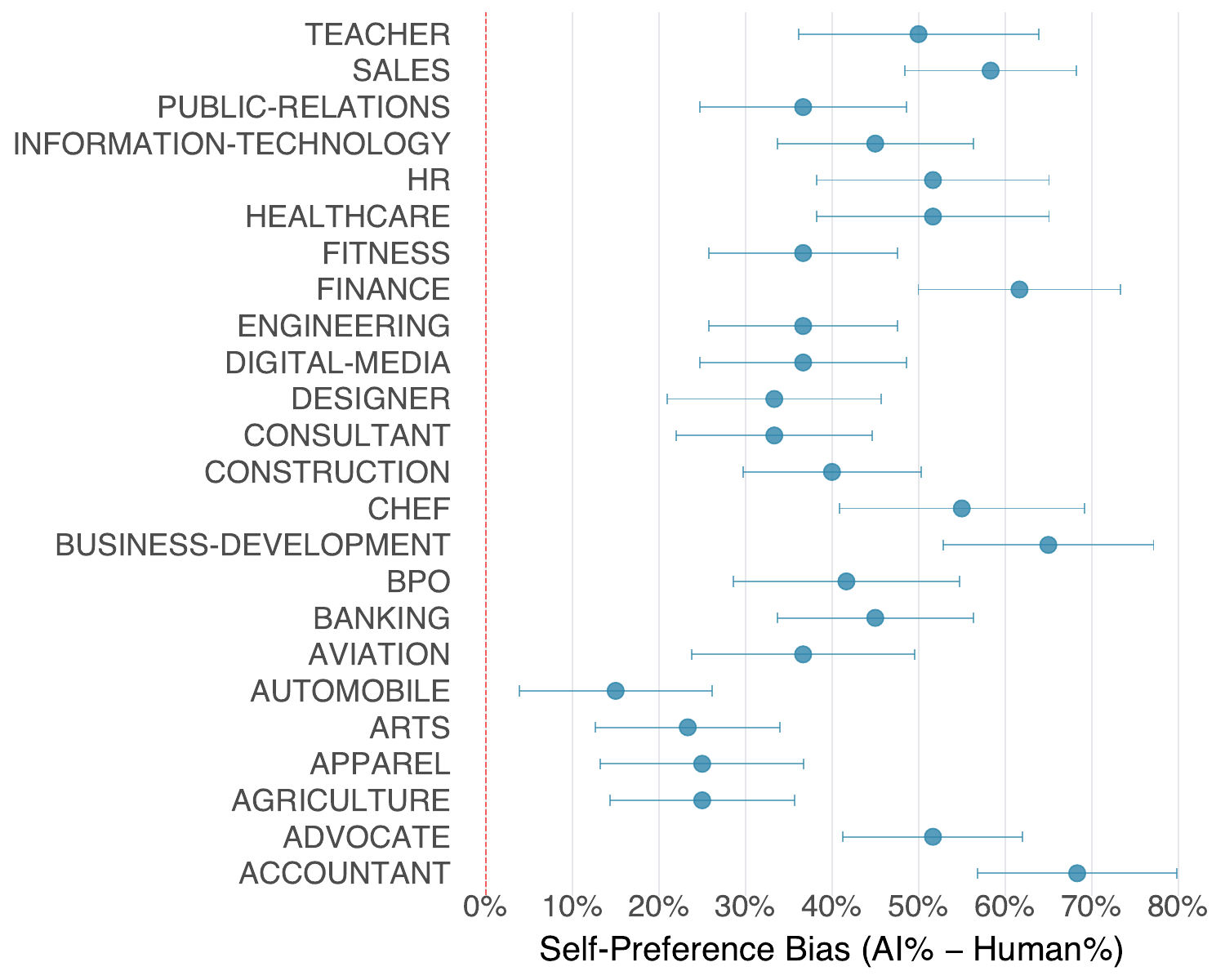}
   \caption{Self-Preference by Job Category Under DeepSeek-V3} 
    \label{fig:bias_category_deepseek}
    \begin{minipage}{\linewidth}
\vspace{1ex}
\footnotesize \textit{Notes:} Each bar shows the self-preference bias across under DeepSeek-V3 in simulated hiring pipelines. Positive values indicate that candidates using the evaluator LLM are more likely to be shortlisted than those submitting human-written resumes. Across occupations, evaluator-generated resumes are consistently overrepresented among those selected.
\end{minipage}
\end{figure}

\begin{figure}[tb]
    \centering
        \centering
        \includegraphics[width=\linewidth]{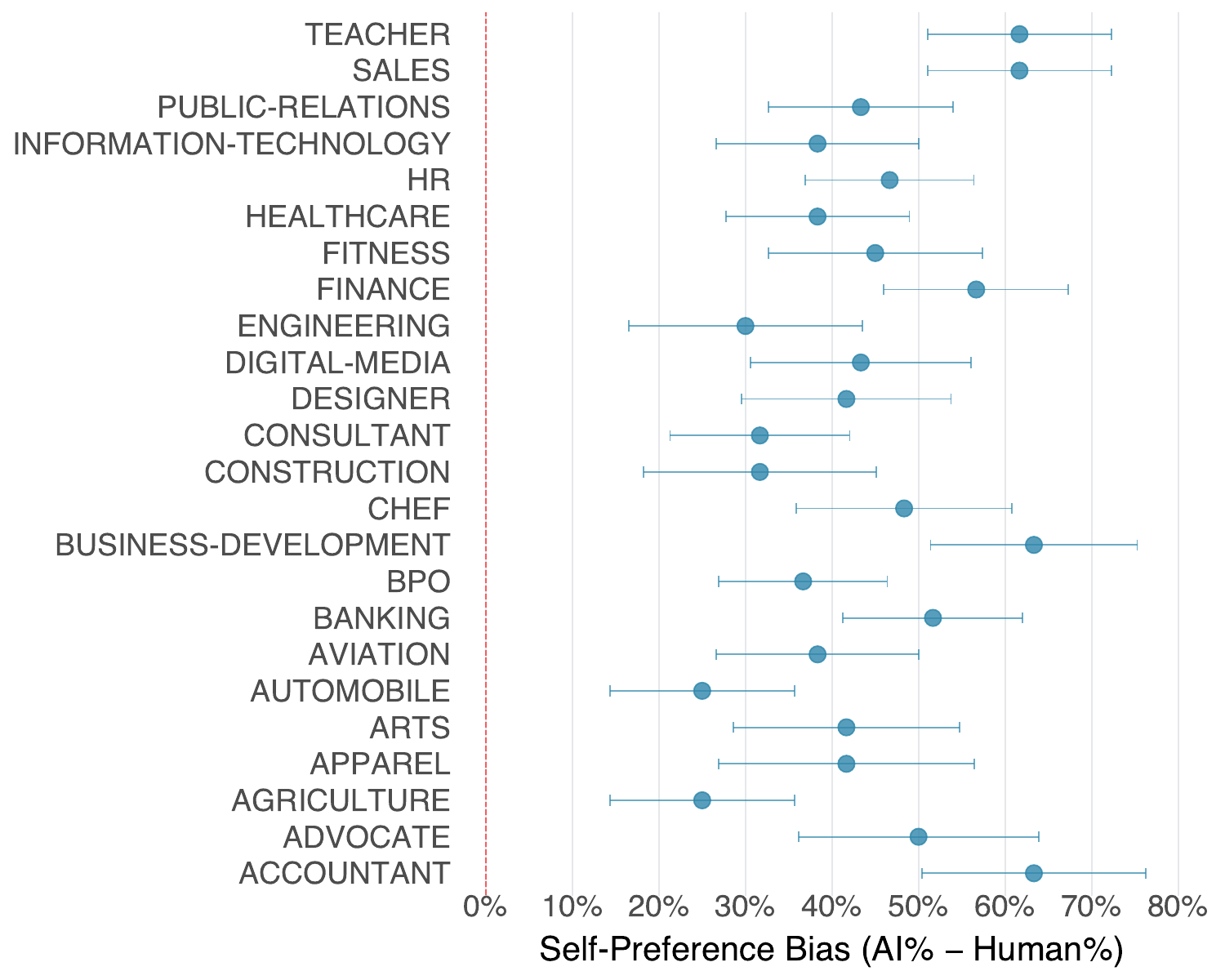}
   \caption{Self-Preference by Job Category Under GPT-4o} 
    \label{fig:bias_category_gpt4o}
    \begin{minipage}{\linewidth}
\vspace{1ex}
\footnotesize \textit{Notes:} Each bar shows the self-preference bias across under GPT-4o in simulated hiring pipelines. Positive values indicate that candidates using the evaluator LLM are more likely to be shortlisted than those submitting human-written resumes. Across occupations, evaluator-generated resumes are consistently overrepresented among those selected.
\end{minipage}
\end{figure}

\begin{figure}[tb]
    \centering
        \centering
        \includegraphics[width=\linewidth]{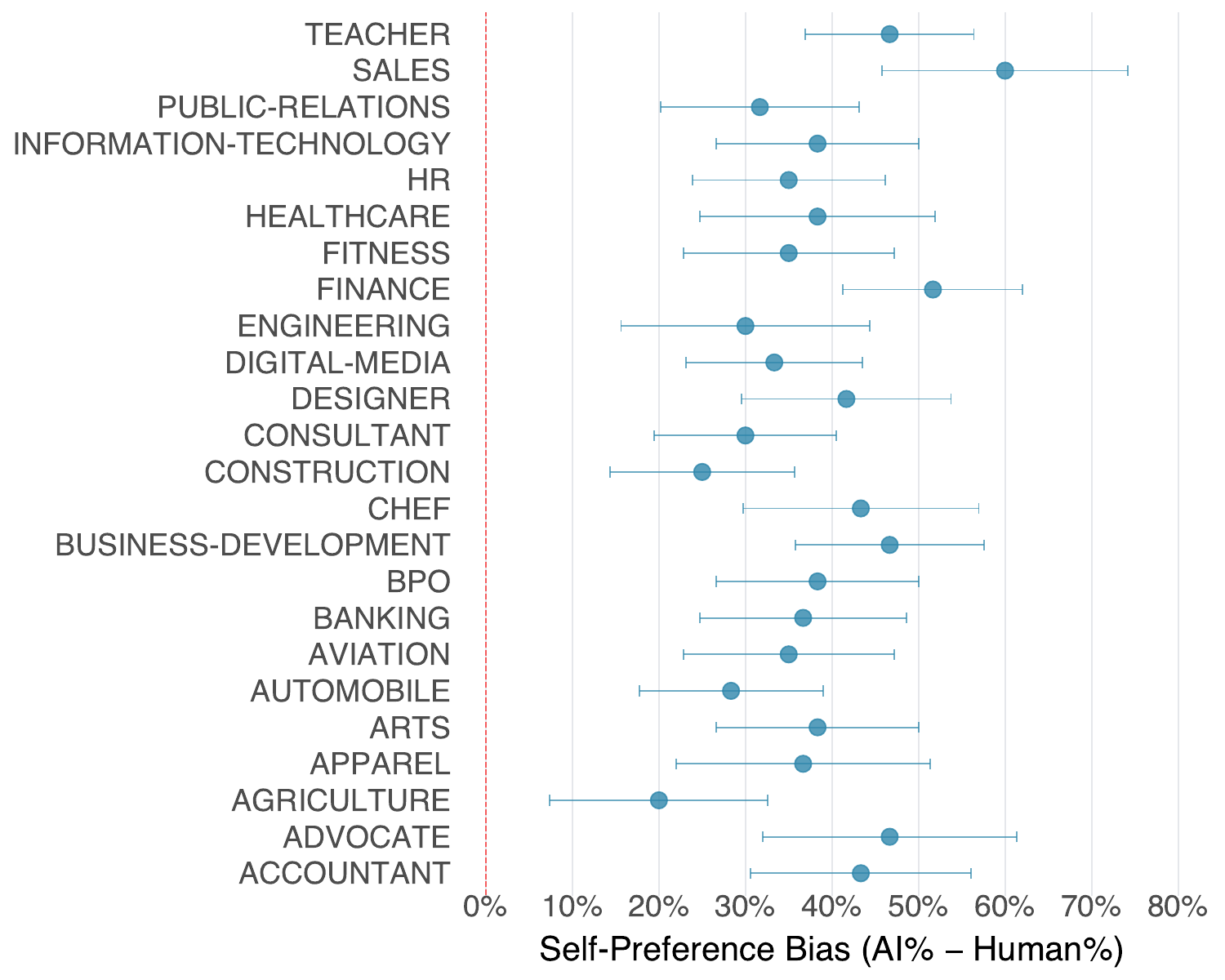}
   \caption{Self-Preference by Job Category Under LLaMA 3.3-70B} 
    \label{fig:bias_category_llam3}
    \begin{minipage}{\linewidth}
\vspace{1ex}
\footnotesize \textit{Notes:} Each bar shows the self-preference bias across under LLaMA 3.3-70B in simulated hiring pipelines. Positive values indicate that candidates using the evaluator LLM are more likely to be shortlisted than those submitting human-written resumes. Across occupations, evaluator-generated resumes are consistently overrepresented among those selected.
\end{minipage}
\end{figure}

\end{APPENDICES}




%
%






\end{document}